\documentclass[aps,prd,reprint,groupedaddress]{revtex4-2}
\usepackage[english]{babel}
\usepackage{microtype}
\usepackage{graphicx}
\usepackage{amsmath,amssymb, graphics, setspace,cancel,multirow}
\usepackage{commath}
\usepackage{mathtools}
\usepackage{color}
\usepackage{soul}
\usepackage{xcolor,cancel}
\usepackage[export]{adjustbox}
\newcommand{\mathsym}[1]{{}}
\newcommand{\unicode}[1]{{}}

\renewcommand{\thesection}{\arabic{section}}
\renewcommand{\thesubsection}{\thesection.\arabic{subsection}}
\renewcommand{\thesubsubsection}{\thesubsection.\arabic{subsubsection}}
\renewcommand{\theparagraph}{\thesubsubsection.\arabic{paragraph}}
\renewcommand{\thesubparagraph}{\theparagraph.\arabic{subparagraph}}

\newcommand{\ORCIDiD}[1]{\href{https://orcid.org/#1}{\includegraphics[width=2ex]{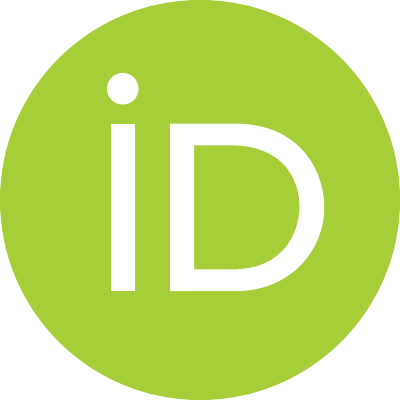}}}

\makeatletter
\def\p@subsection{}
\makeatother

\usepackage[hidelinks,draft=false]{hyperref}

\begin{document}

\title{Estimate of force noise from electrostatic patch potentials in LISA Pathfinder}

\def\unitn{Department of Physics, University of Trento, and Trento Institute for Fundamental Physics and Applications, INFN,  38123 Povo, Trento, Italy}

\author{S~Vitale~\ORCIDiD{0000-0002-2427-8918}}\email{stefano.vitale@unitn.it}\affiliation{\unitn}
\author{V~Ferroni~\ORCIDiD{0000-0002-2260-6658}}\affiliation{\unitn}
\author{L~Sala~\ORCIDiD{0000-0002-2682-8274}}\affiliation{\unitn}
\author{W\,J~Weber~\ORCIDiD{0000-0003-1536-2410}}\affiliation{\unitn}

\date{August 30, 2024}
\begin{abstract}
This paper discusses force noise in LISA and LISA Pathfinder arising from the interaction of patch potentials on the test mass and surrounding electrode housing surfaces with their own temporal fluctuations.  We aim to estimate the contribution of this phenomenon to the force noise detected in LISA Pathfinder in excess of the background from Brownian motion. We introduce a model  that approximates the interacting test mass and housing surfaces as concentric spheres, treating patch potentials as isotropic stochastic Gaussian processes on the surface of these spheres. We find that a scenario of patches due to surface contamination, with diffusion driven density fluctuations, could indeed produce force noise with the observed
frequency $f^{-2}$ dependence. However, there is not enough experimental evidence, neither from LISA Pathfinder itself, nor from other experiments, to predict the amplitude of such a noise, which could range from completely negligible to explaining the entire noise excess. We briefly discuss several measures to ensure that this noise is sufficiently small in LISA.
\end{abstract}

\maketitle

\section{\label{sec:intro} Introduction}

Electrostatic patch potentials are among the  possible sources of force disturbance in LISA and LISA Pathfinder (LPF) \cite{Vitale}. In particular they may be a candidate source to explain part of the excess noise observed in LPF, above the Brownian noise level \cite{Ultimate}, in the frequency band $0.02-1\text{ mHz}$. 

Forces from the stray electrostatic fields produced by patch potentials, have been a leading driver for both the design and operations of the system of the test mass (TM) and surrounding electrode housing (EH), known collectively as the gravitational reference sensor (GRS).  This includes, for instance, the adoption of relatively large (2.9-4~mm) TM/EH gaps \cite{Spie}, the avoidance of applied DC voltages in the electrostatic actuation system \cite{armano2024nanonewton}, and the in-flight measurement and compensation of the averaged stray field that couples TM charge into force \cite{prlelectro,Charge2017}.  Our quantitative modelling and experimental understanding of the relevant stray potentials and the resulting forces are quite good for stray potentials over relatively \textit{macroscopic} scales; force measurement upon controlled modulation of TM charge or electrode potentials allows measurement of the relevant stray potentials averaged over, respectively, the size of the TM face ($\approx 2000$~mm$^2$) or that of an electrode ($\approx 500$~mm$^2$).  As such,  coupling of patch potentials, both steady and noisy, to TM charge and actuation voltages is well modelled and projected to the LPF and LISA performance \cite{prlelectro,Charge2017,armano2024nanonewton,Fromlab}.  

The effect of variations of patch fields on \textit{microscopic} levels, smaller than the electrodes or even gap dimensions, is much less constrained by experiment.  Force noise from the ``patch-to-patch'' interactions on smaller scales are independent of TM charge or averaged electrode potentials, and thus both inaccessible in the modulation experiments mentioned above and always present.  Quantifying the force noise contribution from the interaction of the quasi-static patch fields with their own temporal fluctuations requires modelling, in addition to experimental probes, and this is the subject of this current paper.

In the literature  the problem  has been  analysed either within the model of  two parallel infinite conducting planes, covered by a layer of dipoles and separated by a gap  \cite{Speake96},  or within one  in which a set of random shaped plane capacitors,  formed by tiles of the surface of the cubic shaped LISA TM, face the  equivalent tiles on the inner surface of the EH surrounding the TM \cite{prlelectro}.

The first of the two models above allows   a quantitative parametrization of the patch potential distribution, but lacks the essential topology of a finite TM fully enclosed within a cavity. This is particularly relevant in the case of LISA, where gaps between the TM and the EH may be as large as 10\% of the TM size, so that the infinite plane model may seriously lack in representativeness.

The second model is limited in its assignment of equipotential domains and by the inaccuracy of approximations in the relevant capacitance derivatives for domains that are smaller than the gap separating test mass from electrode housing.

In the attempt of overcoming these limitations, we investigate a model wherein the TM and the EH are  two concentric spheres separated by a vacuum  gap. The shape of the TM and EH  is not the actual one, but the basic topology is, thus the model may be considered as a topological  improvement of the infinite planes one.

The spherical geometry  allows  taking advantage of the expansion of the electrostatic potentials into spherical harmonics, in analogy  with the  potential Fourier expansion of~\cite{Speake96}. The expansion in spherical harmonics has been discussed in \cite{Buchman_2011}, though with an aim and an approach significantly different from ours \footnote{{The problem of GPB  \cite{Buchman_2011} was the coupling of quasi-static patch potentials on the surface of their spherical gyroscopes, with the quasi-static patch potentials on the gyroscope housing. With slow modulations caused by spacecraft motion around the gyros, this coupling exerted torques and caused gyroscope spurious precession that limited the accuracy of the measurements. Patch potential fluctuations were not a limiting factor for the experiment.}}. We will thus work out the necessary equations from first principles. 

Our spherical model also presents limitations in describing the electrostatic interaction with a cubic test mass.  For instance, patch fields on a spherical TM produce no torque sensitivity to the total TM charge, while the same is not true for a cubic TM, with a  non-zero torque sensitivity which  in fact has been directly measured \cite{prlelectro,armano2024nanonewton,Pollack}.  Similarly, the spherical geometry does not allow simple implementation of electrodes with variable voltages, which would allow simulation of force and torque experiments with modulated voltages.  However, we will show that the concentric sphere model does allow to draw useful, quantitative conclusions on the patch-to-patch force noise. Future studies with finite element modelling will be useful to further anchor these conclusions to the realistic cubic geometry.

The paper is structured as follows: in section ~\ref{sect:model} we describe the concentric sphere model and its key parameters. In section~\ref{sec:harmonic} we summarise the spherical harmonic formalism for the calculation of the electrostatic potential within our model. In section~\ref{sec:forceterms} we use these results to calculate the force on the inner sphere, i.e. on the TM, as a function of the amplitude of the spherical harmonic components. In section~\ref{sect:distribution} we parametrise the statistical properties of such spherical harmonic amplitudes, and calculate the statistical properties of the force noise. In section~\ref{sect:LPF} we apply the derived framework to the analysis of a series of experiments related to  patch potentials, performed on LPF by means of a properly applied TM charge bias. In Sect.~\ref{sect:proj} we finally use our model, and the information from LPF, and other relevant experiments reported in the literature, to evaluate the force noise that may have been caused by the patch potential fluctuations in LPF. Section~\ref{sect:conc} gives a few concluding remarks, especially related to the impact of our findings for the LISA mission.

\section {\label{sect:model} The concentric spheres model}
\begin{figure}[h]
\begin{center}
\includegraphics[width=1.1\columnwidth]{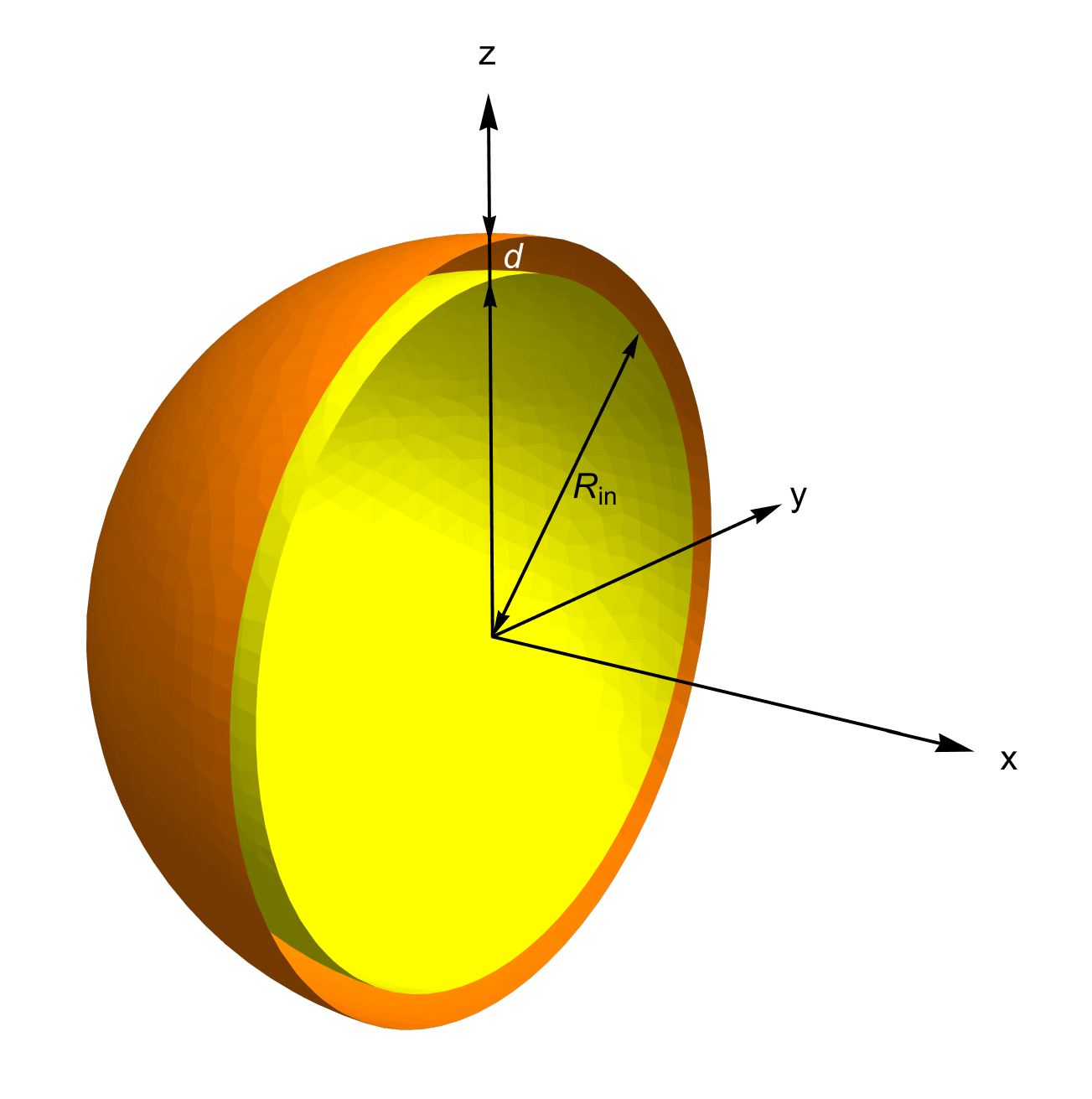}
\end{center}
\caption{Schematics of the test-mass - electrode housing system: two concentric spherical surfaces of radius $R_{in}$ and $R_{out}=R_{in}+d$ respectively.}
\label{fig:scheme}
\end{figure}

In the  model we discuss here,  we treat the TM outer surface  and the EH inner surface, as two concentric, spherical conducting surfaces of radius $R_{in}$ and $R_{out}$ respectively (see Fig. \ref{fig:scheme}). Both surfaces are assumed to be  covered by a  continuous  layer of dipoles, each dipole being radially oriented.  Finally, the inner surface  is electrically floating while the outer is grounded.

When performing numerical calculations we adopt an inner sphere with the same area as the cubic LISA test mass  $R_{in}=L\sqrt{6/(4 \pi)}\simeq 32\text{ mm}$, with $L=46 \text{ mm}$ the side-length of the LISA cubic TM. We will also take $R_{out}= R_{in}+d=35.5\text{ mm}$, with $d= 3.5\text{ mm}$ an approximate average, over the three faces of the TM, of the gaps between TM and EH in LISA.

\section{\label{sec:harmonic} Spherical harmonic analysis of electrostatic potential}
\subsection{\label{sec:potential} Potential of a spherical surface dipole distribution.}
The expansion of the Green function for the Laplace equation, in \emph{real} spherical harmonics %(see the Appendix \ref{sec:sph}) 
of spherical coordinates $r$, $\theta$, and $\phi$, is:
\begin{widetext}
\begin{equation}
\label{eq:green}
\begin{split}
&G\left(r_1,\theta_1,\phi_1,r_2,\theta_2,\phi_2\right)=\frac{1}{\sqrt{r _1^2+r _2^2-2r _2 r _1 \left(\sin \left(\theta _1\right) \sin \left(\theta _2\right) \cos \left(\phi _1-\phi _2\right)+\cos \left(\theta _1\right) \cos \left(\theta _2\right)\right)}}=\\
&= \sum_{l=0}^\infty \frac{4\pi}{2l+1} \sum_{m=-l}^{l}  \frac{r_{{\scriptscriptstyle<}}^l }{r_{\scriptscriptstyle>}^{l+1} } Y^{-m}_l(\theta_1, \phi_1) Y^m_l(\theta_2, \phi_2)
\end{split}
\end{equation}
\end{widetext}
where $r_{{\scriptscriptstyle<}}$ is the smallest between $r_1$ and $r_2$, and $r_{{\scriptscriptstyle>}}$ the largest.

If a surface charge distribution $\sigma\left(\theta,\phi\right)$ covers the surface of a sphere of radius $R$, the electrical potential $V\left(r,\theta,\phi\right)$ at the point of spherical coordinates $r$, $\theta$, and $\phi$ is:

\begin{equation}
\label{eq:potential}
\begin{split}
&V \left(r,\theta,\phi \right)=\\
&= \frac{1}{ \varepsilon_0 R}\sum_{l=0}^\infty \frac{1}{2l+1} \sum_{m=-l}^{l}  \left(\frac{R }{r} \right)^{l+1}Y^{m}_l(\theta, \phi) c_{l,m}\text{    for   } r >R\\
&= \frac{1}{ \varepsilon_0 R}\sum_{l=0}^\infty \frac{1}{2l+1} \sum_{m=-l}^{l}\left( \frac{r}{R}\right)^{l}  Y^{m}_l(\theta, \phi) c_{lm}\text{    for   } r <R\\
\end{split}
\end{equation}
with
\begin{equation}
\label{eq:coef}
c_{l,m}=R^2\int_{0}^\pi\int_0^{2\pi}Y^m_l(\theta', \phi')\sigma\left(\theta',\phi'\right)\sin{\theta'}d\theta' d\phi',
\end{equation}
the spherical harmonic amplitudes of $R^2\sigma(\theta,\phi)$.

Spatially varying potential on the surface of conductors have been described by a distribution of electrostatic dipoles normal to the surface, with surface density $D$ \cite{Jackson}.  For a conducting sphere, we can treat this as two surface charge distributions  of opposite polarity, $\sigma_+(\theta,\phi)$ and $\sigma_-(\theta,\phi)$ on two spheres of radius $R+\delta/2$ and $R-\delta/2$ respectively, and then take the limit for $\delta\to0$. As the total charge must be zero, we assume that $\sigma_+(\theta,\phi)=\left(R^2/(R+\delta/2)^2\right)\sigma(\theta,\phi)$ and $\sigma_-(\theta,\phi)=-\left(R^2/(R-\delta/2)^2\right)\sigma(\theta,\phi)$.

Then the potential inside the inner sphere is 
\begin{equation}
\label{eq:dipoleout}
\begin{split}
&V \left(r,\theta,\phi \right)=\\
&=\lim_{\delta\to 0} \frac{1}{ \varepsilon_0}\sum_{l=0}^\infty \frac{1}{2l+1} \sum_{m=-l}^{l}  r^l  Y^{-m}_l(\theta, \phi) c_{lm}\times\\&\times\left(\frac{1}{\left(R+\delta/2\right)^{l+1} }-\frac{1}{\left(R-\delta/2\right)^{l+1} }\right)=\\
&=- \frac{1}{ \varepsilon_0 R^2}\sum_{l=0}^\infty \frac{l+1}{2l+1} \sum_{m=-l}^{l} \left(\frac{r}{R}\right)^l  Y^{m}_l(\theta, \phi) b_{l,m}
\end{split}
\end{equation}
with the dipole \emph{harmonic amplitudes} $b_{l,m}$ defined by:
\begin{equation}
\label{eq:coefdip}
\begin{split}
&b_{l,m}=\\&\lim_{\delta\to 0}R^2\int_{0}^\pi\int_0^{2\pi}Y^m_l(\theta', \phi')\left(\sigma\left(\theta',\phi'\right)\times \delta\right)\sin(\theta')d\theta' d\phi '\equiv \\
&\equiv R^2\int_{0}^\pi\int_0^{2\pi}Y^m_l(\theta', \phi' )D\left(\theta',\phi'\right)\sin(\theta')d\theta' d\phi'
\end{split}
\end{equation}
which defines   the surface dipole density as 
\begin{equation}
\label{eq:surfacedensity}
D\left(\theta,\phi\right)=\lim_{\delta\to 0}\left( \sigma\left(\theta,\phi\right)\times \delta\right)=\sum_{l=0}^\infty\sum_{m=-l}^l \frac{b_{l,m}}{R^2}Y_l^m(\theta,\phi)
\end{equation}

The formula could have been obtained by deriving the second line of Eq.~\eqref{eq:potential} relative to $R$. A similar calculation for the potential outside the sphere gives
\begin{equation}
\label{eq:dipolein}
\begin{split}
&V \left(r,\theta,\phi \right)=\\
&=  \frac{1}{ \varepsilon_0R^2}\sum_{l=0}^\infty \frac{l}{2l+1} \sum_{m=-l}^{l} \left(\frac{R}{r}\right)^{l+1} Y^{m}_l(\theta, \phi) b_{l,m}
\end{split}
\end{equation}

Eqs. \ref{eq:dipoleout} and \ref{eq:dipolein} show that the potential has a discontinuity at $r=R$ given by
\begin{equation}
\label{eq:disc}
\Delta V\left(\theta,\phi \right)=\frac{1}{ \varepsilon_0 R^{2}}\sum_{l=0}^\infty \sum_{m=-l}^{l}   Y^{m}_l(\theta, \phi) b_{l,m}=\frac{D\left(\theta,\phi \right)}{\varepsilon_0}
\end{equation}

\subsection{\label{sec:image} Images}
If the sphere, on which the dipole distribution lies, is conducting, there will be an induced surface charge that can be calculated by the method of images. Table~\ref{tab:image} summarises the key features of the images of a point charge, and of a point radial dipole, in the presence of a conducting spherical surface of radius $R$.
\begin{table}[h]
\caption{Charge and dipole images on a grounded conducting sphere of radius $R$ of a point charge and a point dipole with radial orientation. Relations hold  for both $R>r$ and $R<r$. For a neutral sphere, instead of a grounded one,  and $R<r$, the total charge of images is neutralised by an opposite point charge in the center.}
\begin{equation*}
\begin{array}{|c|c|c|c|c|c|}
\hline
\multicolumn{3}{|c|}{\text{Source}} & \multicolumn{3}{|c|}{\text{Image} }\\
\hline
\text{radial}&\text{charge}&\text{dipole}&\text{radial}&\text{charge}&\text{dipole}\\
\hline
\text{position}&\text{}&\text{}&\text{position}&\text{}&\text{}\\
\hline
r&q&-&R^2/r&-q R/r&-\\
\hline
r&-&p&R^2/r&p R/r^2&p R^3/r^3\\
\hline
\end{array}
\end{equation*}
\label{tab:image}
\end{table}

Now assume that our radially aligned dipole, of dipole moment $p$, lies in between our two concentric spheres at a position $R_s$, with  $R_{in}\le R_s\le R_{out}$. The dipole will produce images on both spheres, images  that will be in turn reflected by the spheres. We calculate with Mathematica the  sequence in Table~\ref{tab:gapimage}.

\begin{table}[h]
\caption{Positions and intensities of images of a radial dipole of moment $p$ sitting at position $R_s$ between two conducting spheres of radius $R_{in}$ and $R_{out}$ respectively.}
\begin{equation*}
\arraycolsep=2 pt\def\arraystretch{2.}
\begin{array}{|c|c|c|c|c|c|}
\hline
\multicolumn{3}{|c|}{\text{Inner images}}&\multicolumn{3}{|c|}{\text{Outer images}}\\
\hline
\text{Position}&\text{Dipole}&\text{Charge}&\text{Position}&\text{Dipole}&\text{Charge}\\
\hline
\frac{R_{in}^2}{R_s} & \frac{p R_{in}^3}{R_s^3} & \frac{p R_{in}}{R_s^2} & \frac{R_{out}^2}{R_s} & \frac{p R_{out}^3}{R_s^3} & \frac{p R_{out}}{R_s^2} \\
 \frac{R_{in}^2 R_s}{R_{out}^2} & \frac{p R_{in}^3}{R_{out}^3} & 0 & \frac{R_{out}^2 R_s}{R_{in}^2} & \frac{p R_{out}^3}{R_{in}^3} & 0 \\
 \frac{R_{in}^4}{R_{out}^2 R_s} & \frac{p R_{in}^6}{R_{out}^3 R_s^3} & \frac{p R_{in}^2}{R_{out} R_s^2} & \frac{R_{out}^4}{R_{in}^2 R_s} & \frac{p R_{out}^6}{R_{in}^3 R_s^3} & \frac{p R_{out}^2}{R_{in} R_s^2} \\
 \frac{R_{in}^4 R_s}{R_{out}^4} & \frac{p R_{in}^6}{R_{out}^6} & 0 & \frac{R_{out}^4 R_s}{R_{in}^4} & \frac{p R_{out}^6}{R_{in}^6} & 0 \\
 \frac{R_{in}^6}{R_{out}^4 R_s} & \frac{p R_{in}^9}{R_{out}^6 R_s^3} & \frac{p R_{in}^3}{R_{out}^2 R_s^2} & \frac{R_{out}^6}{R_{in}^4 R_s} & \frac{p R_{out}^9}{R_{in}^6 R_s^3} & \frac{p R_{out}^3}{R_{in}^2 R_s^2} \\
 \frac{R_{in}^6 R_s}{R_{out}^6} & \frac{p R_{in}^9}{R_{out}^9} & 0 & \frac{R_{out}^6 R_s}{R_{in}^6} & \frac{p R_{out}^9}{R_{in}^9} & 0 \\
 \frac{R_{in}^8}{R_{out}^6 R_s} & \frac{p R_{in}^{12}}{R_{out}^9 R_s^3} & \frac{p R_{in}^4}{R_{out}^3 R_s^2} & \frac{R_{out}^8}{R_{in}^6 R_s} & \frac{p R_{out}^{12}}{R_{in}^9 R_s^3} & \frac{p R_{out}^4}{R_{in}^3 R_s^2} \\
 \frac{R_{in}^8 R_s}{R_{out}^8} & \frac{p R_{in}^{12}}{R_{out}^{12}} & 0 & \frac{R_{out}^8 R_s}{R_{in}^8} & \frac{p R_{out}^{12}}{R_{in}^{12}} & 0 \\
 \hline
\vdots&\vdots&\vdots&\vdots&\vdots&\vdots\\
 \hline
\end{array}
\end{equation*}
\label{tab:gapimage}
\end{table}
\subsection{Potential due to images}
Take now a surface distribution of radial dipoles on a sphere of radius $R_s$, with surface density given by Eq.~\eqref{eq:surfacedensity}, with $R=R_s$.
Each elementary dipole of intensity $p=\frac{b_{l,m}}{R_s^2}Y_l^m(\theta,\phi) R_s^2  \sin{\theta} d\phi d\theta$, at position $\theta$ and $\phi$ on the sphere will have its own sequence of images with intensity and radial positions given in Table~\ref{tab:gapimage}, and the same angular position as the elementary dipole.

As a consequence the entire surface distribution will produce images consisting of surface distribution of charge and dipoles, with the same angular dependence of the original distribution, with amplitudes scaled from the original by the same factor by which  $p$ is scaled in  Table~\ref{tab:gapimage}, and finally lying on a sphere with radius given again by Table~\ref{tab:gapimage}.

Note that, in principle, one should neutralise the inner sphere by adding a charge opposite to the total charge of the images. However, for $l>0$, $\int_0^\pi\int_0^{2\pi}Y_l^m(\theta,\phi)\sin(\theta)d\theta d\phi=0$, and carry no net charge. 

As for the component with $l=0$, that is a spherically symmetric dipole density, it requires no image, and its only effect is to change the potential of the inner, electrically floating, conducting surface by a constant, without inducing any surface charge on either of the spheres. Its presence is irrelevant for force calculations so that, from now on, all sums will start from $l=1$.

To the original distribution, and to all its images, one can apply the formulas of Sect.~\ref{sec:potential}, and calculate the resulting potential. The angular dependence will be the same for all images, while the radial dependence will produce a term proportional either to $r^l$ or to $r^{-(l+1)}$. 

These can be grouped together for all images. We obtain, with the help of  Mathematica:
\begin{equation}
\label{eq:impot}
\begin{split}
&V_I(r,\theta ,\phi )=\frac{1}{R_{in}^2 \varepsilon_0}\sum _{l=1}^{\infty } \frac{1}{2 l+1}\sum _{m=-l}^l  b_{l,m}Y_l^m(\theta ,\phi )\left(\frac{R_{in}}{R_s}\right)^{l+2}\times\\
&\times   \left(\frac{ \left((l+1) R_{in}^{2 l+1}+l R_s^{2 l+1}\right)}{R_{in}^{2 l+1}-R_{out}^{2 l+1}}\left(\frac{r}{R_{in}}\right)^l-\right.\\&\left.-\frac{ \left((l+1) R_{out}^{2 l+1}+l R_s^{2 l+1}\right)}{R_{in}^{2 l+1}-R_{out}^{2 l+1}}\left(\frac{R_{in}}{r}\right)^{l+1}\right)
\end{split}
\end{equation}

One can check that
\begin{equation}
\begin{split}
\label{eq:limrin}
&\lim_{r\to R_{in}}V_I(r,\theta ,\phi )=\\&=\frac{1}{R_{s}^2 \varepsilon_0}\sum _{l=1}^{\infty } \frac{1}{2 l+1}\sum _{m=-l}^l  b_{l,m} Y_l^m(\theta ,\phi )(l+1)  \left(\frac{R_{in}}{R_s}\right)^l
\end{split}
\end{equation}
which is the opposite of the potential created by the original distribution at $r=R_{in}$, while, equivalently,
\begin{equation}
\begin{split}
&\lim_{r\to R_{out}}V_I(r,\theta ,\phi )=\\&=-\frac{1}{R_{s}^2 \varepsilon_0}\sum _{l=1}^{\infty } \frac{1}{2 l+1}\sum _{m=-l}^l  b_{l,m} Y_l^m(\theta ,\phi )l  \left(\frac{R_{s}}{R_{out}}\right)^{l+1}
\end{split}
\end{equation}
again the opposite of the  potential created by the original distribution at $r=R_{out}$.

\subsection{Total potential for the actual configuration}

We will now consider two surface distributions of dipoles, one with $R_s=R_{in}$ and one with $R_s=R_{out}$, with coefficients $b_{l,m}^{in}$ and $b_{l,m}^{out}$ respectively, with aim of calculating the force they produce on the inner sphere. To calculate the force we need two potentials:
\begin{itemize}
\item The total potential generated by all sources on the surface of the inner conducting sphere. This is needed to calculate the total electric field at the conductor surface, and, from that, the resulting surface charge density. 
Notice that the potential due to the dipoles on the inner sphere must be calculated by taking $R_s>R_{in}$ and only at the end of any calculation take the limit $R_s \to R_{in}$.
\item The potential due to sources on the outer sphere only. These include the dipoles lying on the inner surface of the outer sphere, and the images located at ${r}>R_{out}$. This is needed to calculate the field and its gradient that exert forces on the dipoles and the charges lying on the surface of the inner sphere.
\end{itemize}

The first term, the total potential at $r\simeq R_{in}$  can be derived from Eqs.~\eqref{eq:dipoleout}, \eqref{eq:dipolein}, and \eqref{eq:impot} to be
\begin{equation}
\label{eq:totpot}
\begin{split}
&V_{tot}(r,\theta ,\phi )=\frac{1}{R_{in}^2 \varepsilon_0}\sum _{l=1}^{\infty } \sum _{m=-l}^l Y_l^m(\theta ,\phi )\times\\
&\times\left( \left(\frac{\alpha ^{2 l+1}}{1-\alpha ^{2 l+1}}+\frac{l+1}{2 l+1}\right) b_{l,m}^{in}+\frac{\alpha ^{l+2} }{1-\alpha ^{2 l+1}} b_{l,m}^{out}\right)\times\\&\times\left(\left(\frac{R_{in}}{r}\right)^{l+1}-\left(\frac{r}{R_{in}}\right)^l\right)
\end{split}
\end{equation}
with $\alpha=R_{in}/R_{out}\simeq0.90$, while the second is
\begin{equation}
\begin{split}
\label{eq:extpot}
&V_{out}(r,\theta ,\phi )=-\frac{1}{R_{\text{in}}^2 \varepsilon_0}\sum _{l=1}^{\infty } \sum _{m=-l}^l Y_l^m(\theta ,\phi )\times\\
& \left(\frac{\alpha ^{2 l+1} }{1-\alpha ^{2 l+1}}b_{l,m}^{in}+\frac{\alpha ^{l+2}}{1-\alpha ^{2 l+1}} b_{l,m}^{out}\right)\left(\frac{r}{R_{in}}\right)^l
\end{split}
\end{equation}

{Note that at $r=R_{in}$, i.e. on the conducting surface of the  inner sphere, the potential in Eq.~\eqref{eq:totpot} is zero, but  its gradient, which is what we need to calculate the  surface charge density, is not.} 

Before concluding this section it is useful to establish a connection between the formulas above, and the potentials one would measure just outside the dipole distribution lying on the surface of the inner sphere, or just inside the surface of the outer sphere, the real `patch potentials'.

For the former, as the potential due to the dipoles on the outer sphere is independently cancelled by images at the inner sphere, one needs to evaluate the contribution of the dipoles on the inner sphere only.
 
This can be obtained by adding the potential in Eq.~\eqref{eq:limrin}, evaluated for $R_s\to R_{in}$ and $b_{l,m}=b_{l,m}^{in}$, to that in Eq.~\eqref{eq:dipolein}, evaluated for $R \to R_{in}$, $r\to R_{in}$, and $b_{l,m}=b_{l,m}^{in}$. Such  calculation gives:
\begin{equation}
\label{eq:tmonly}
\begin{split}
&V_K(\theta ,\phi )=\frac{1}{R_{in}^2 \varepsilon_0}\sum _{l=1}^{\infty }\sum _{m=-l}^l  b_{l,m}^{in}Y_l^m(\theta ,\phi )=\frac{D^{in}(\theta,\phi)}{\varepsilon_0}
\end{split}
\end{equation}
with $D_{in}(\theta,\phi)$ the surface density of dipoles on the inner sphere. This result is consistent with the starting assumption that patch potentials can be described as a surface distribution of dipoles.

Note that the result is independent of $R_{out}$, and thus that would be the potential also in the case  $R_{out}\to \infty$, that is in the case of an isolated floating sphere covered by a dipole distribution. This may represent the case of an isolated sample one would investigate with a Kelvin probe, to detect the spatial variation of the potential, as done in Ref.~\cite{Robertson_2006}. This is the reason for the suffix $K$ in Eq.~\eqref{eq:tmonly}

A similar calculation gives for the potential just inside the outer sphere:
\begin{equation}
\label{eq:dV}
\begin{split}
&V'_K(\theta ,\phi )=-\frac{1}{R_{out}^2 \varepsilon_0}\sum _{l=1}^{\infty }\sum _{m=-l}^l  b_{l,m}^{out}Y_l^m(\theta ,\phi )=-\frac{D^{out}(\theta,\phi)}{\varepsilon_0}
\end{split}
\end{equation}
with $D^{out}(\theta,\phi)$ the surface dipole density on  the outer sphere.

One can check that the solution of the Laplace equation for the potential in the gap between the spheres, with the boundary conditions expressed by Eqs.~\eqref{eq:tmonly} and~\eqref{eq:dV}, matches that obtained by summing up the potential due to the images and that due to the original dipole distributions, a useful test of consistency for our calculations.

\section{\label{sec:forceterms} Field, gradient and force on the inner sphere}

\subsection{General formula for the force}

To evaluate the component of the force on the inner sphere along some direction, we calculate the effect of the field and field gradient due to sources on the outer sphere, on the dipole and charge distributions lying on the surface of the inner sphere. 

The alternative approach, preferred in planar geometries, of calculating the derivative of the electrostatic energy with respect to a displacement of the inner sphere along the selected direction leads to cumbersome formulas, as such a displacement would break the spherical symmetry.

As the system is statistically isotropic, we can pick any of the three Cartesian components. In spherical coordinates, calculations are significantly easier for the z component that we calculate as:

\begin{equation}
\label{eq:force}
\begin{split}
&F_{z,c}=R_{in}^2\int_{0}^\pi \int_0^{2\pi} \left(D^{in}(\theta,\phi)\partial_r E_{out,z}(\theta,\phi)+\right.\\&\left.+ \sigma(\theta,\phi)E_{out,z}(\theta,\phi)\right) \sin(\theta)d \theta d \phi
\end{split}
\end{equation}

In Eq.~\eqref{eq:force} we have used:
\begin{itemize}
\item{the surface density of dipoles sitting on the inner sphere, the explicit expansion of which we repeat here for convenience:
\begin{equation}
\label{eq:A}
D^{in}(\theta,\phi)=\sum_{l=1}^\infty\sum_{m=-l}^l\frac{b_{l,m}^{in}}{R_{{in}}^2}Y_l^m(\theta,\phi);
\end{equation}}
\item{the surface charge due to images on the inner sphere
\begin{equation}
\label{eq:B}
\sigma(\theta,\phi)=\varepsilon_0 E_{tot,r}(\theta,\phi) =-\varepsilon_0\left(\frac{\partial V_{tot}(r,\theta,\phi)}{\partial r}\right)_{r=R_{in}};\end{equation} }
\item{the z component of the field created, at the surface of the inner sphere, by the dipoles and images  on the outer sphere:
\begin{equation}
\label{eq:C}
\begin{split}
&E_{out,z}(\theta,\phi)=-\hat{z}\cdot\left(\vec{\nabla}V_{out}(r,\theta,\phi)\right)_{r=R_{in}}=\\
&=-\left(\cos(\theta)\frac{\partial V_{out}(r,\theta,\phi)}{\partial r}-\right.\\&\left.\sin(\theta)\frac{1}{r}\frac{\partial V_{out}(r,\theta,\phi)}{\partial \theta}\right)_{r=R_{in}};
\end{split}
\end{equation}
}
\item{and, finally, the radial derivative of the above
\begin{equation}
\label{eq:D}
\begin{split}
&\partial_r E_{out,z}(\theta,\phi)=\\&=-\left(\cos(\theta)\frac{\partial^2 V_{out}(r,\theta,\phi)}{\partial r^2}-\right.\\&\left.-\sin(\theta)\frac{\partial}{\partial r}\left(\frac{1}{r }\frac{\partial V_{out}(r,\theta,\phi)}{\partial \theta}\right)\right)_{r=R_{in}}.
\end{split}
\end{equation}}
\end{itemize}
\subsection{Charge density from images}
Differentiating the potential in Eq.~\eqref{eq:totpot} { relative to $r$}, and following Eq.~\eqref{eq:B} and \ref{eq:C}, we obtain:
\begin{equation}
\label{eq:rho}
\begin{split}
&\sigma(\theta,\phi)=\frac{1}{R_{in}^3 }\sum _{l=1}^{\infty } \sum _{m=-l}^l Y_l^m(\theta ,\phi )\times\\&\times\frac{\left(l \alpha ^{2 l+1}+l+1\right) b_{l,m}^{\text{in}}+(2 l+1) \alpha ^{l+2} b_{l,m}^{\text{out}}}{1-\alpha ^{2 l+1}}
\end{split}
\end{equation}
\subsection{Field and its radial derivative}
As for $E_{out,z}(\theta,\phi)$ one can check with Mathematica that
\begin{equation}
\begin{split}
&\cos(\theta)\frac{\partial r^l Y_l^m(\theta,\phi)}{\partial r}-\frac{\sin(\theta)}{r}\frac{\partial r^l Y_l^m(\theta,\phi)}{\partial \theta}=\\&=r^{l-1}\sqrt{\frac{2 l+1}{2 l-1}} \sqrt{l^2-m^2} Y_{l-1}^m(\theta ,\phi )
\end{split}
\end{equation}

From this, it follows that
\begin{equation}
\label{eq:e}
\begin{split}
&E_{out,z}(\theta,\phi)=\frac{1}{R_{in}^3 \varepsilon_0}\sum _{l=1}^{\infty } \sum _{m=-(l-1)}^{l-1}\left(\frac{\alpha ^{2 l+1} }{1-\alpha ^{2 l+1}}b_{l,m}^{in}+\right.\\&\left.+\frac{\alpha ^{l+2}}{1-\alpha ^{2 l+1}} b_{l,m}^{out}\right)\sqrt{\frac{2 l+1}{2 l-1}} \sqrt{l^2-m^2} Y_{l-1}^m(\theta ,\phi ),
\end{split}
\end{equation}
and that
\begin{equation}
\label{eq:dre}
\begin{split}
&\partial_r E_{out,z}(\theta,\phi)=\frac{1}{R_{in}^4 \varepsilon_0}\sum _{l=1}^{\infty } \sum _{m=-(l-1)}^{l-1} \left(\frac{\alpha ^{2 l+1}}{1-\alpha ^{2 l+1}} b_{l,m}^{in}+\right.\\&\left.+\frac{\alpha ^{l+2}}{1-\alpha ^{2 l+1}}b_{l,m}^{out}\right)\sqrt{\frac{2 l+1}{2 l-1}} \sqrt{l^2-m^2} (l-1)Y_{l-1}^m(\theta ,\phi )
\end{split}
\end{equation}

Note that in the sum in Eq.~\eqref{eq:dre}, the term with $l=1$ is zero due to the multiplication by $l-1$. Thus the first term contains $Y_1^m$ and would exert no force on the $Y_0^0$ component of the dipole surface density of the inner sphere. This confirms the legitimacy of ignoring the term with $l=0$ in all calculations.
\subsection{Final formula for the force}
One can now evaluate the integral in  Eq.~\eqref{eq:force}. To do that we will first consider that  $Y_l^m=0$ for both $|m|>l$ and $l=-1$, so that in Eqs.~\eqref{eq:e} and \eqref{eq:dre} it is better to make the substitution $l \to l+1$. Using then the orthonormality of spherical harmonics one gets 
\begin{equation}
\label{eq:forcefinal}
\begin{split}
&F_z=\\&=\frac{1}{R_{\text{in}}^4 \varepsilon_0}\sum _{l=1}^{\infty } \sum _{m=-l}^{l}\frac{ \sqrt{(2 l+3)(2 l+1)((l+1)^2-m^2)}}{\left(1-\alpha ^{2 l+1}\right) \left(1-\alpha ^{2 l+3}\right)}\alpha^{3+l}\times\\
&\times\left(\alpha^{l}b_{l,m}^{in} b_{l+1,m}^{in} +b_{l,m}^{in} b_{l+1,m}^{out} +\right.\\&\left.+\alpha^{2 l+2}b_{l,m}^{out} b_{l+1,m}^{in} +\alpha^{l+2}b_{l,m}^{out} b_{l+1,m}^{out} \right)
\end{split}
\end{equation}

To describe patch potential fluctuations, we will assume in the following that the dipole densities are time-dependent and that the harmonic amplitudes are then functions of time.

\section{\label{sect:distribution} Statistical distribution of harmonic amplitudes,  and autocorrelation of the force}
We need now to discuss the possible distributions  of  the harmonic amplitudes  $b_{l,m}^{in}$ or $b_{l,m}^{out}$   that would reproduce a corresponding realistic distribution of patches. We will discuss first the static part of those amplitudes.
\subsection{Isotropic random fields on a sphere}
The idea of distributions consisting of spherical harmonics with random amplitudes is not new. One can find a vast literature on \emph{isotropic Gaussian random fields on the sphere} (see for instance \cite{spherical}). These are random fields defined over the surface of a unit sphere,  for which, among other things, the autocorrelation function at two points on the sphere, with coordinates $\lbrace \theta_1, \phi_1\rbrace$ and $\lbrace \theta_2, \phi_2\rbrace$ respectively, is just a function of the spherical angle $\omega=\arccos\left(\cos(\theta_1)\cos(\theta_2)+\sin(\theta_1)\sin(\theta_2)\cos(\phi_1-\phi_2)\right)$ between the two points.

As discussed  in \cite{spherical}, if our surface dipole density is one of these processes,  the autocorrelation can be expanded as:
\begin{equation}
\label{eq:autocorrelation}
\begin{split}
&\langle D\left(\theta_1,\phi_1\right) D\left(\theta_2,\phi_2\right)\rangle=\\&=\frac{1}{R^4}\sum_{l=0}^\infty \frac{2l+1}{4\pi}A_l \text{P}_l\left(\cos(\omega)\right)=\\&=\frac{1}{R^4}\sum_{l=0}^\infty\sum_{m=-l}^lA_lY_l^m(\theta_1,\phi_1)Y_l^m(\theta_2,\phi_2).
\end{split}
\end{equation}

The second equality is based on the well known expansion of Legendre polynomials of the cosine of the spherical angle, into complex spherical harmonics of the two spherical positions, but it can be easily shown to hold also for real spherical harmonics.

One can then check, by using Eq.~\eqref{eq:coefdip}, that  $\langle b_{l,m} b_{j,n}\rangle=A_l \delta_{l,j}\delta_{m,n}$, thus the  $A_l$'s are positive numbers that play the role of a power spectral density (PSD) as a function of the `spherical frequency' $l$. 

Note that the division by $R^4$ on the right-hand side of the first line, and on the second line of Eq.~\eqref{eq:autocorrelation}, is necessary to obtain such a direct relation between $\langle b_{l,m}^2\rangle$ and $A_l $.

If the dipole surface density is time varying but stationary, in the sense of stochastic processes, then Eq.~\eqref{eq:autocorrelation} becomes

\begin{equation}
\label{eq:autocorrelationtime}
\begin{split}
&\mathcal{R}_D(\omega,\Delta t)\equiv\langle D\left(\theta_1,\phi_1,t\right) D\left(\theta_2,\phi_2,t+\Delta t\right)\rangle=\\&=\frac{1}{R^4}\sum_{l=0}^\infty \frac{2l+1}{4\pi}A_l(\Delta t) \text{P}_l\left(\cos(\omega)\right)
\end{split}
\end{equation}

From Eq.~\eqref{eq:autocorrelationtime} one can easily calculate that, as $P_l(1)=1$,
\begin{equation}
\label{eq:sigrms1}
\begin{split}
&\langle D(\theta,\phi,t) D(\theta,\phi,t+\Delta t)\rangle=\frac{1}{R^4}\sum _{l=0}^{\infty }\frac{2l+1}{4 \pi}A_l(\Delta t)
\end{split}
\end{equation}

The cross coherence of the two time-domain, statistically identical stochastic processes $D\left(\theta_1,\phi_1,t\right)$ and  $D\left(\theta_2,\phi_2,t\right)$ is then given by:
\begin{equation}
\label{eq:cohere}
\begin{split}
&\frac{\langle D(\theta,\phi,t) D(\theta,\phi,t+\Delta t)\rangle}{\sqrt{\langle D^2(\theta,\phi,t)\rangle\langle D^2(\theta,\phi,t+\Delta t)\rangle}}=\\&=\frac{\sum_{l=0}^\infty \frac{2l+1}{4\pi}A_l(\Delta t) \text{P}_l\left(\cos(\omega)\right)}{\sum_{l=0}^\infty \frac{2l+1}{4\pi}A_l(\Delta t) }
\end{split}
\end{equation}

Eq.~\eqref{eq:autocorrelationtime} allows us to fix the scale of the $A_l(0)$ by linking it to the mean square fluctuation, at any given time, of patch potentials over the surface,  {a quantity that could be measured} with a Kelvin probe. This, provided that the measurement is faster than the typical time scale on which $A(l)$ changes significantly, or in general, if the time variations can be neglected. 

Indeed, by using Eq.~\eqref{eq:tmonly}, one gets
\begin{equation}
\label{eq:rmsstat}
\begin{split}
& V_{K,rms}^2\equiv \langle V_K(\theta,\phi,t)^2\rangle=\frac{1}{\varepsilon_0^2}\langle {D^{in}(\theta,\phi,t)}^2\rangle=\\&=\frac{1}{\varepsilon_0^2 R_{in}^4 }\sum _{l=0}^{\infty }\frac{2l+1}{4 \pi}A_l(0)
\end{split}
\end{equation}

{ These relations between voltage fluctuations and the $A_l(0)$ suggest the definition $V_l^2(\Delta t)\equiv A_l(\Delta t)/(\varepsilon_0^2 R_{in}^4)$, which makes many of the formulas more transparent.}

\subsection{Autocorrelation of force}
We now calculate the autocorrelation of force on the inner sphere, assuming that both the inner and outer dipole distribution are stationary and isotropic stochastic processes with{, in addition,} $b_{0,0}^{in}(t)=b_{0,0}^{out}(t)=0$, $\langle b_{l,m}^{in}(t)\rangle=\langle b_{l,m}^{out}(t)\rangle=0$, $\langle b_{l,m}^{in}(t)b_{j,n}^{in}(t+\Delta t)\rangle=\alpha^4\langle b_{l,m}^{out}(t)b_{j,n}^{out}(t+\Delta t)\rangle=A_l(\Delta t)\delta_{l,j}\delta_{m,n}$, and finally $\langle b_{l,m}^{in}(t)b_{j,n}^{out}(t+\Delta t)\rangle=0$. The correction by radius ratio $\alpha=R_{in}/R_{out}$ is needed to give the same dipole surface density to both spheres.

\begin{figure}[htbp]
\includegraphics[width=\columnwidth,center]{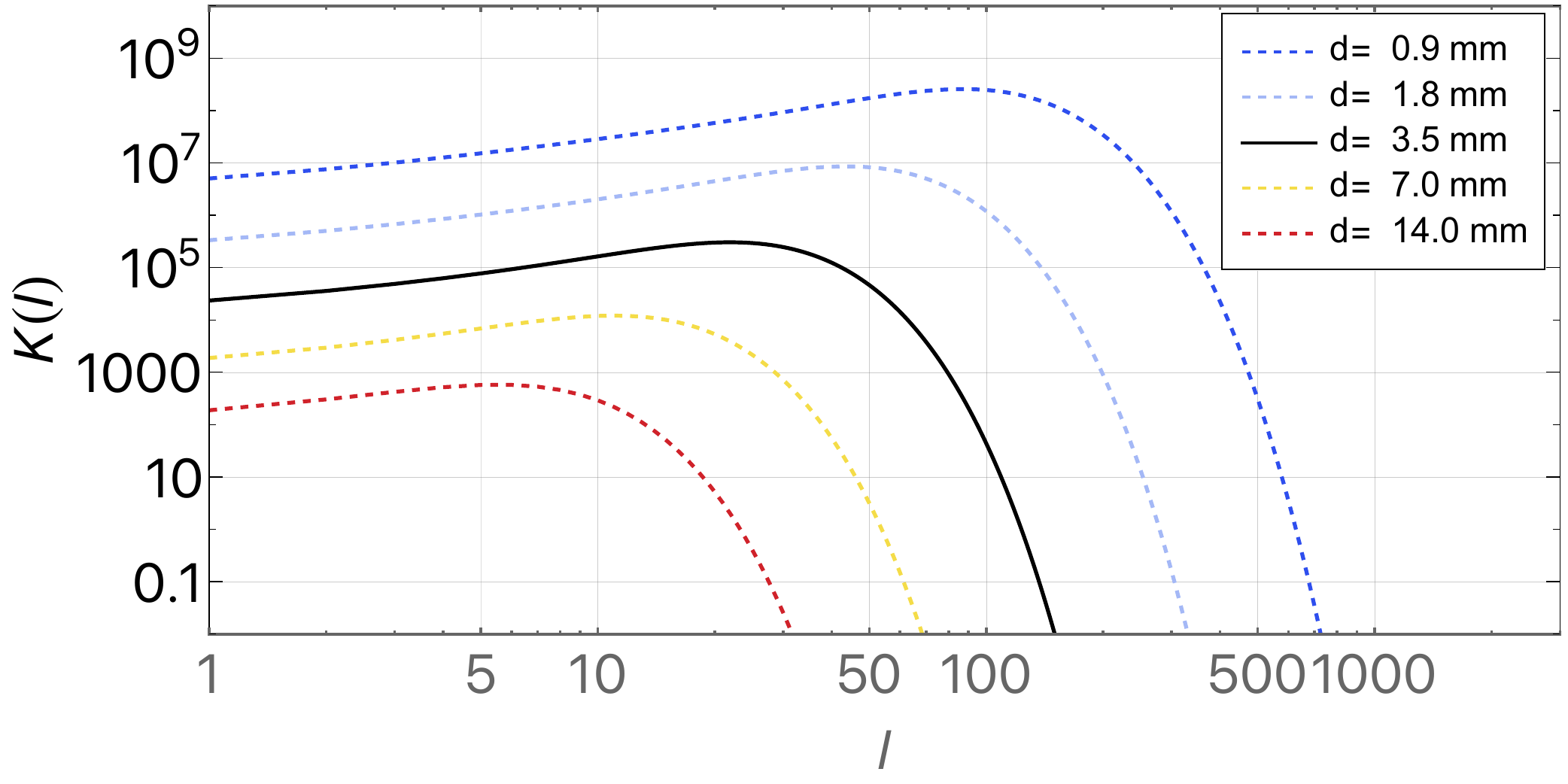}
\caption{The transfer factor $K(l)$ defined in Eq.~\eqref{eq:kappa}, as a function of the harmonic number $l$. Black continuous line: the case for the gap value $d=3.5 \text{ mm}$ we are using in all calculations. Dashed coloured lines: the case for both wider and narrower gaps as shown in the legend.}
\label{fig:kappa}
\end{figure}

With these prescriptions, as $\langle b_{l,m}^{in}\rangle=\langle b_{l,m}^{out}\rangle=0$, and $\langle b_{l,m}^{in}b_{l+1,m}^{in}\rangle=0$, and same for the outer dipoles, then $\langle F_z\rangle=0$. The force has instead a non zero autocorrelation given by: 
\begin{equation}
\label{eq:forceautocorr}
\begin{split}
&\mathcal{R}_F(\Delta t)\equiv\langle F_z(t)F_z(t+\Delta t)\rangle=\\&\frac{1}{R_{\text{in}}^8 \varepsilon_0^2}\sum _{l=1}^{\infty }K(l)A_l(\Delta t)A_{l+1}(\Delta t)=\\
&=\varepsilon_0^2\sum _{l=1}^{\infty }K(l)V_l^2(\Delta t)V_{l+1}^2(\Delta t)
\end{split}
\end{equation}
with
\begin{equation}
\begin{split}
&K(l)=(2l+1)(2l+3)\left(\sum_{m=-l}^l\left((l+1)^2-m^2\right)\right)\times\\
&\times\left(\frac{\alpha^{l+3}}{(1-\alpha^{2l+1})(1-\alpha^{2l+3})}\right)^2\times\\
&\times\left((\alpha^{l})^2+\alpha^{-4}\left((\alpha^{2l+2})^2+1\right)+\alpha^{-8}(\alpha^{l+2})^2\right)
\end{split}
\end{equation}
which gives
 \begin{equation}
\label{eq:kappa}
\begin{split}
&K(l)=\frac{1}{3} (l+1) (4 l (l+2)+3)^2\frac{\alpha ^{2 l+2} \left(1+\alpha ^{2 l}\right) \left(1+\alpha ^{2 l+4}\right)}{\left(1-\alpha ^{2 l+1}\right)^2 \left(1-\alpha ^{2 l+3}\right)^2}
\end{split}
\end{equation}
Before picking a spectral shape for $A_l$ and calculating $\langle F_z(t)F_z(t+\Delta t)\rangle$, it is interesting to look at the dependence on $l$ of the `transfer' factor $K(l)$, which we show in Fig.~\ref{fig:kappa}, for the value of the gap between the spheres,  $d=3.5 \text{ mm}$, we are using here, but also, for the sake of comparison, for narrower and wider gaps.

For all values of the gap, $K(l)$ reaches a maximum,  and decays quite rapidly after that. By defining $l_{100}$, the value of $l$  for which $K(l)$ attains $\simeq 1\%$ of such a maximum, we find that, within, at worst, a 10\% accuracy, $l_{100}\simeq 2.4 \pi R_{in}/d$, with $l_{100}=72$ in our case of $d=3.5\text{ mm}$.

Thus when the `wavelength' $\pi R_{in}/(l/2)$ of the $l$-th harmonic component, becomes as large as $\simeq 1.2\,d$, the force effect is already strongly suppressed.  Crudely speaking, this is because smaller wavelengths correspond to smaller patches, and the electric field lines due to patches smaller than the gap close on the adjacent patches on the same sphere rather than crossing the gap and closing somewhere on the other sphere. This obviously suppresses the electrical interaction between the spheres.

Anyway this result shows that under realistic hypotheses, numerical calculations can be safely truncated at $l$ a few times $l_{100}$.

{The possible dependency of $V_l^2(\Delta t)$ on $\Delta t$ will be discussed later. For now, we note that if  $V_l^2(\Delta t)$, in the above range of $l$ less than a few hundreds, follows a simple exponential form, $V_l^2(\Delta t) \propto e^{-\lvert \Delta t \rvert/\tau_l}$, where $\tau_l$ is a correlation time constant, and if $\tau_l \gtrsim 10^5 \text{ s}$,  then for frequencies $f \ge 0.02\text{ mHz}$ the Fourier transform of $\mathcal{R}_F(\Delta t)$, which is one half of  the power spectral density (PSD) of the force , is $\propto 1/f^2$. Consequently, the amplitude spectral density (ASD), being the square root of the PSD, is $\propto 1/f$. This would match the frequency dependency of the excess force noise ASD observed in LPF \cite{subfemtog}.}

As a final note, the force is bilinear in the stochastic processes $b_{l,m}^{in}(t)$ and $b_{l,m}^{out}(t)$. So, even if these are Gaussian, the force in principle is not. However the force results from the incoherent sum of many processes, and thus, unless just one or two dominate its fluctuation, the central limit theorem should guarantee that it may be considered Gaussian for  the purposes of the present work.

% The  experiments on patch potentials we have mentioned above \cite{Charge2017} \cite{prlelectro} \cite{Robertson_2006}, report values for a dc part of the patch potentials and/or of the force they induce.

% It is important to realise that instead,  within the model we are discussing here, as both $\langle V_K(t)\rangle=0$ and $\langle F_z(t)\rangle=0$, there is strictly speaking no static part of both the potential and the force.

% However a measurement of 
% \begin{equation}
% % \label{eq:timeav}
% \overline{F}_z(t)=\frac{1}{T}\int_t^{t+T} F_z(t) dt
% \end{equation}
% though still with zero mean value, has autocorrelation
% \begin{equation}
% % \label{eq:favaut}
% \begin{split}
% &\langle\overline{F}_z(t)\overline{F}_z(t+\Delta t)\rangle=\frac{1}{T^2}\int_t^{t+T}\int_{t+\Delta t}^{t+\Delta t+T}\langle F_z(t)F_z(t')\rangle dt\;dt'=\\
% &=\frac{1}{T}\int_{-T}^T\mathcal{R}_F\left(\Delta t'-\Delta t\right) \left(1-\frac{\lvert \Delta t'\rvert}{T}\right)d\Delta t'
% \end{split}
% \end{equation}

% Thus a single sample will be a random number from a distribution with a width $D_{\overline{F}}=\sqrt{\langle\overline{F}_z(t)\overline{F}_z(t)\rangle}$, varying over a time scale set by the evolution of $\mathcal{R}_F\left(\Delta t\right)$ with $\Delta t$.

% Such a picture is not {inconsistent} with the experimental evidence that has often witnessed very long term evolution of the results of nominally static measurements \cite{Robertson_2006} \cite{Charge2017}. We will discuss this aspect again when examining the details of the LPF experiment.
%%%%%
\section{\label{sect:LPF}LPF experiments with charge bias, and the implications for \texorpdfstring{$A_l(\Delta t)$}{Al(Dt)}}
% \section{\label{sect:LPF}LPF experiments with charge bias, and the implications for $A_l(\Delta t)$}

Let us now discuss how the   LPF measurements  on patch potentials can be imported in our model, with the aim of using their results to derive $V_l^2(\Delta t)$. 

\subsection{\label{sec:Q} Charge bias in the concentric sphere model and in LPF}

In LPF, patch potentials were investigated by applying a static charge bias $Q$ to the TM and by measuring both the resulting static force, and the resulting increase of the force noise \cite{Charge2017}. 

In our model, a charge $Q$ on the TM creates a uniform charge density on its surface $\sigma_Q=Q/(4 \pi R_{in}^2)={Q} Y_0^0(\theta,\phi)2 \sqrt{\pi}/(4 \pi R_{in}^2)$. This surface density couples to $E_{z,out}$ in Eq.~\eqref{eq:e}, to create a force:
\begin{equation}
\label{eq:forceQ}
\begin{split}
&F_{Q,z}(t)=Q\sqrt{ \frac{3}{4 \pi}} \frac{1}{R_{in}^3 \varepsilon_0}\frac{\alpha ^3 }{1-\alpha ^3} \left(b_{1,0}^{in}(t)+ b_{1,0}^{out}(t)\right)
\end{split}
\end{equation}

To compare this formula with that used in the relevant LPF literature, \cite{armano2024nanonewton,Charge2017,prlelectro}, it is useful to define the effective, rescaled potential difference $\Delta V_{K,1}(t)=\frac{b_{1,0}^{in}(t)+ b_{1,0}^{out}(t)}{\varepsilon_0 R_{in}^2}$. {Using the assumptions on the statistical properties of $b_{1,0}^{in}(t)$ and of $b_{1,0}^{out}(t)$, the autocorrelation of $\Delta V_{K,1}(t)$ becomes} $\langle \Delta V_{K,1}(t)\Delta V_{K,1}(t+\Delta t)\rangle=\frac{1+\alpha^4}{\alpha^4}V_1^2(\Delta t)$. With such a definition:
\begin{equation}
\label{eq:forcedx}
\begin{split}
&F_{Q,z}(t)=\frac{Q\Delta V_{K,1}(t)}{R_{in}}\sqrt{ \frac{3}{4 \pi}}\frac{\alpha^3 }{1-\alpha ^3}
\end{split}
\end{equation}
%%%%%

In LPF, the coupling $dF_{Q,z}/dQ$ was measured by introducing a step change in the TM charge, through the UV discharge system \cite{Charge2017}, and then applying the appropriate compensation voltage to the four identical electrodes facing, in pairs, the two opposite TM x-faces, to null the effect.

The resulting compensation force scales as  $d F_{Q,z}^c/dQ=-Q \Delta_x(1/C_T)(\partial C_X/\partial x)$, where $C_T$ is the total capacitance of the TM to ground, $C_X$ is that of one of the electrodes relative to the TM, and where $\Delta_x$  is the sum of the bias voltages applied to the electrode pair on one x-face, minus the sum of the voltages applied to the opposite pair.

Equating this equation to Eq.~\eqref{eq:forcedx} we get

\begin{equation}
\label{eq:deltax}
\begin{split}
&\Delta V_{K,1}(t)=\frac{R_{in}}{\sqrt{ \frac{3}{4 \pi}}\frac{\alpha^3 }{1-\alpha ^3}}\frac{1}{C_T}\frac{\partial C_X}{\partial x} \Delta_x(t)\equiv \beta \Delta_x(t)
\end{split}
\end{equation}
with $\beta\simeq0.2$.

By using  Eq.~\eqref{eq:deltax}, we may use the measurement of $\Delta_x$ from  \cite{armano2024nanonewton,Charge2017}, to infer the properties of $V_{1}^2(\Delta t)$. { In particular $V_{1}^2(\Delta t)$ is connected to the autocorrelation $\mathcal{R}_{\Delta_x}(\Delta t)=\left\langle \Delta _x(t) \Delta _x(\text{$\Delta $t}+t)\right\rangle$ of $\Delta_x(t)$ via:
\begin{equation}
\label{eq:aonepre}
\begin{split}
V_1^2(\Delta t)&=\frac{\alpha^4}{1+\alpha^4}\langle \Delta V_{K,1}(t)\Delta V_{K,1}(t+\Delta t)\rangle=\\
&=\beta^2\frac{\alpha^4 }{\alpha ^4+1}\mathcal{R}_{\Delta_x}(\Delta t)\\
\end{split}
\end{equation}
}

% {\subsection{LPF charge bias experiment: quasi static force, noise increase  and the estimate of $V_1^2(\Delta t)$}
{\subsection{LPF charge bias experiment: quasi static force, noise increase  and the estimate of \texorpdfstring{$V_1^2(\Delta t)$}{V12(Dt)}}

LPF  measured the quasi-static value of $\Delta_x$ for both its TMs \cite{LPF}, that we indicate as TM1 and TM2, at several points in time  over the $\simeq15$ months of the mission  operations \cite{Charge2017} (see Table~\ref{tab:deltax}).
\begin{table}[h]
\caption{Measurements of $\Delta_x(t)$ for both TM1 ($\Delta_{x,1}(t)$) and TM2 ($\Delta_{x,2}(t)$) of LPF as a function of time $t$ since LPF launch. The data on rows 1 to 4 are from Ref.~\cite{Charge2017}, while the data on  row 5 have been obtained only  after the publication of that paper.
}
$
\begin{array}{|c|c|c|c|}
\hline
\text{n}&t \text{(d)}&\Delta_{x,1}(t) (\text{mV}) &\Delta_{x,2}(t) (\text{mV})\\
\hline
 1&110 & -21.02\pm 0.07 & -0.70\pm 0.07 \\
2& 155 & -20.93\pm 0.04 & 0.70\pm 0.03 \\
 3&383 & -18.3\pm 0.9 & \text{not available} \\
4& 395 & -18.7\pm 1.4 & -0.10\pm 0.20 \\
 5&540 & -18.\pm 9. & 2.70\pm 0.05 \\
 \hline
\end{array}
$
% \hline
\label{tab:deltax}
\end{table}

As mentioned in Sect.~\ref{sec:Q}, these  measurement  were performed with controlled steps in the TM charge. $\Delta_x$ on each of the steps was  measured by averaging over $T\simeq 10^4\text{ s}$ \cite{Charge2017}.  

As shown in Table~\ref{tab:deltax}, the measured values displayed variations of several mV for each TM, around average values observed to be $\simeq-20$ mV for TM1 and $\simeq+1$ mV for  TM2.

Similar long term variations of the average quasi-static potentials on centimeter scale samples have been consistently observed in all experiments on patch potentials  \cite{Robertson_2006,Pollack,prlelectro}. They have also been observed to be  sensitive to large variations of the  vacuum environment \cite{Robertson_2006}, something that  is commonly taken as an indication that patch potentials, in these kinds of systems, are dominated by the deposition of  volatile surface contaminants.

In connection with this last point, we note that the average pressure around TM1 and TM2 in LPF, measured from the Brownian noise level, decreased by a factor of $\simeq 10$ over the time span of the measurements in Table~\ref{tab:deltax} \cite{armano2024}. 
The slow change observed in $\Delta_{x,1}(t)$ and $\Delta_{x,2}(t)$ and the long term decrease in pressure may indeed be related,  by the decrease in  surface contamination through outgassing.  Drawing further conclusions from this qualitative observation would be rather speculative.

Before proceeding further  we need to clarify one point. Within the model we are discussing here, as $\langle V_K(t)\rangle=0$ there is strictly speaking no static part  in any of the random quantities we are dealing with here, including $\Delta_x(t)$. However, our averaged measurement:
\begin{equation}
% \label{eq:timeav}
\overline{\Delta}_x(t)=\frac{1}{T}\int_t^{t+T} \Delta_x(t') dt'
\end{equation}
though still with zero mean value, has auto-correlation
\begin{equation}
% \label{eq:favaut}
\begin{split}
&{\langle\overline{\Delta}_x(t)\overline{\Delta}_x(t+\Delta t)\rangle}=\\&=\frac{1}{T^2}\int_t^{t+T}\int_{t+\Delta t}^{t+\Delta t+T}\langle \Delta_x(t)\Delta_x(t')\rangle dt\;dt'=\\
&=\frac{1}{T}\int_{-T}^T\mathcal{R}_{\Delta_x}\left(\Delta t'-\Delta t\right) \left(1-\frac{\lvert \Delta t'\rvert}{T}\right)d\Delta t'\simeq \mathcal{R}_{\Delta_x}\left(\Delta t\right)
\end{split}
\end{equation}
where the last approximation holds for a $\mathcal{R}_{\Delta_x}(\Delta t)$ decaying on a much longer time than $T$.

In this last scenario of a very slowly decaying auto-correlation,  a single sample of $\overline{\Delta}_x(t)$ will be a random number from a distribution with a width $\sqrt{\mathcal{R}_{\Delta_x}(0)}$, while the relative fluctuation from one sample to another one, measured after a time  $\Delta t$, will be of the order $\sqrt{1-\mathcal{R}_{\Delta_x}(\Delta t)/\mathcal{R}_{\Delta_x}(0)}$. Such a picture seems consistent with the long term variations we were discussing above.

In addition to the quasi-static part of $\Delta_x$, LPF also measured  its in-band \footnote{By `in-band' we mean here within the 0.02-1 mHz frequency range} noise \cite{Charge2017,armano2024nanonewton} through
 the detected increase in the differential force noise amplitude spectral density (ASD)  with a highly charged TM ($\simeq 2 \times 10^8\,e$, or $\simeq 1$ V TM bias) \cite{armano2024nanonewton}.

 The extra noise corresponds to a fluctuation of the difference  $\Delta\Delta_x(t)=\Delta_{x,1}(t)-\Delta_{x,2}(t)$, with an ASD at 0.1 mHz of $S_{\Delta\Delta_x,exp}^{1/2}(f=0.1 \text{ mHz})\simeq (0.15\pm0.02)\text{ mV}/\sqrt{\text{Hz}}$ \cite{armano2024nanonewton}. 

 Assuming similar and independent fluctuations in both $\Delta_{x,1}(t)$ and $\Delta_{x,2}(t)$, the result above  corresponds to an ASD for each of the two of $S_{\Delta_x,exp}^{1/2}(f=0.1 \text{ mHz})\simeq (0.11\pm0.01)\text{ mV}/\sqrt{\text{Hz}}$.

The frequency dependence of the PSD of this  $\Delta_x$ noise  is well fit to a linear combination of $1/f$ and $1/f^2$ power laws, with the $1/f^2$ component being completely negligible except, marginally, around the lowest measured frequency of $\simeq0.05$ mHz \cite{armano2024nanonewton}. 

Thus, in essence, the ASD of such  noise, the square root of the PSD, is $\propto 1/f^{1/2}$, and an  estimate on how large  a residual $\propto 1/f$ contribution could be, a relevant question for our discussion later,  gives  a $2\sigma$ upper limit for a single TM of $S_{\Delta_x}^{1/2}(f)\le 65\;\mu\text{V}/\sqrt{\text{Hz}}(0.1 \text{ mHz}/f)$ \footnote{This figure is taken from the posterior distribution behind the fit to the in-band noise of figure 9 of Ref.~\cite{armano2024nanonewton}.}.

It must be said that a joint analysis of both force and torque coupling to the TM charge is consistent with the hypothesis that this $\Delta_x$ noise is dominated by uncorrelated additive voltage noise on the four x-face electrodes, likely from the actuation electronics. The magnitude of this voltage noise matches the ground test results for such electronics noise \cite{armano2024nanonewton}.

This does not rule out the possibility that a fraction of the noise, roughly up to $10\%$ of the $1/f^{1/2}$ component (based on measurement errors), and all of the $1/f$ component, may be contributed by patch potential fluctuations.

Note that the observation that the quasi-static measurements of $\Delta_x$ are in reality random samples of  slowly varying zero-mean stochastic processes poses the  question of how large the ASD of this noise may be in the 0.02--1 mHz band, and if such a projected ASD is compatible with the direct measurements of the in-band ASD we have just discussed.

We have addressed this question by noting that the hypothesis that the process is Gaussian and zero-mean implies that its autocorrelation determines its entire statistics. In particular it allows to calculate both the joint likelihood of the quasi-static measurements and the ASD of the process.

We have then performed   a Bayesian fit to the data of Table~\ref{tab:deltax} assuming that $\Delta_{x,1}(t)$ and $\Delta_{x,2}(t)$ are Gaussian zero-mean stochastic processes both with the same autocorrelation $\mathcal{R}_{\Delta_x}(\Delta t)$.

We have used  three different models for $\mathcal{R}_{\Delta_x}(\Delta t)$ that, under proper assumptions, correspond to  ASDs that, in the band $f>0.02\text{ mHz}$, depend on the frequency as $\propto 1/f^{1/2}$, $\propto1/f$, and $\propto1/f^2$ respectively.

The details of this Bayesian fit are reported in  Appendix \ref{app:bayes}. Here we just summarize, in Table~\ref{tab:projn}, the noise ASD that these three models project for $f\ge0.02\text{ mHz}$.
\begin{table}[h]
    \centering
    \caption{Projected   ASD of $\Delta_x(t)$, for $f\ge 0.02\text{ mHz}$, from the results of the Bayesian fit}
   $\begin{array}{|c|c|}
   \hline
       \text{Model ASD}&S_{\Delta_x}^{1/2}(f)\text{ mV}/\sqrt{\text{Hz}}  \\
       \hline
        \propto 1/f^{1/2}& 184._{-40.}^{+61.}\times\left(\frac{0.1\text{ mHz}}{f}\right)^{1/2}\\
        \hline
       \propto 1/f & 1.27_{-0.33}^{+0.6}\times\left(\frac{0.1\text{ mHz}}{f}\right)\\
        \hline
        \propto 1/f^2 & 0.45_{-0.14}^{+0.28}\times10^{-3}\times\left(\frac{0.1\text{ mHz}}{f}\right)^2\\
        \hline
    \end{array}$
    \label{tab:projn}
\end{table}

The results in Table~\ref{tab:projn} show that the $\propto 1/f^{1/2}$ and $\propto 1/f$ models project an ASD that is too large and incompatible with the observed one.

For the first model the ASD at 0.1 mHz should be a fraction of $0.1\text{ mV}/\sqrt{\text{Hz}}$ while the projection exceeds $100\text{ mV}/\sqrt{\text{Hz}}$. For the second the ASD at 0.1 mHz should be  $\le 0.07\text{ mV}/\sqrt{\text{Hz}}$ while the projection exceeds $1\text{ mV}/\sqrt{\text{Hz}}$.

Thus only the `reddest' model $\propto1/f^2$ that projects a negligible in-band noise, is compatible with the observations. 

The conclusion we draw from this analysis is that the fluctuations of the quasi-static part of  the patch potentials, within a simple power law model, do not project  any significant in-band noise. 

Thus a useful parametrisation of the LPF  observations, is that the auto-correlation of $\Delta_x(t)$ splits as $\mathcal{R}_{\Delta_x}(\Delta t)=\mathcal{R}_{\Delta_x,dc}(\Delta t)+\mathcal{R}_{\Delta_x,n}(\Delta t)$, with  $\mathcal{R}_{\Delta_x,dc}(\Delta t)$ describing the long term fluctuations of the quasi-static (`$dc$') part of $\Delta_x(t)$, and  $\mathcal{R}_{\Delta_x,n}(\Delta t)$ describing its in-band  noise (`$n$'). 

For the first part one could take the reddest model  discussed above, with $\mathcal{R}_{\Delta_x,dc}(\Delta t)=\sigma_{\Delta_x,dc}^2 f_2(\Delta t,\tau_{dc})$ ($f_2(\Delta t,\tau)$ is defined in Eq.~\eqref{eq:effe2}), and   $\sigma_{\Delta_x,dc}$ and $\tau_{dc}$  equal, respectively, to  $\sigma_{\Delta_x}$ and $\tau$ on the bottom line of Table~\ref{tab:fitp} in Appendix~\ref{app:bayes}. We stress that this is just a useful parametrisation, and that there is no particular physical motivation behind it.

For the in-band noise  we only have  upper limits to the possible $ 1/f^{1/2}$ and $1/f$ components of the ASD, and just in a limited frequency range. This information is insufficient to calculate  $\mathcal{R}_{\Delta_x,n}(\Delta t)$.

However, in view of the possibility that a  $1/f$ component in the patch potentials fluctuation  may have played some role  in  the force noise excess in LPF, we explore the simple exponential model of Eq.~\eqref{eq:effe1}  $\mathcal{R}_{\Delta_x,n}(\Delta t)=\sigma_{\Delta_x,n}^2e^{-\lvert\Delta t\rvert/\tau_n}$ that, for $2 \pi f\tau_n\gg1$, has indeed a $1/f$ ASD (see Eq.~\eqref{eq:lim}). 

To be compatible with the observations $\sigma_{\Delta_x,n}$ and $\tau_n$ have to fulfill the following constraints:
\begin{enumerate}
    \item to get the $1/f$ behavior across the frequency band of interest, $2 \pi f\tau_n\gg1$ for $f\ge 0.02\text{ mHz}$;
    \item to be compatible with the measurements,  $S_{\Delta_x,n}^{1/2}(f)$, the ASD corresponding to $\mathcal{R}_{\Delta_x,n}(\Delta t)$, should obey to $S_{\Delta_x,n}^{1/2}(f)\le  65\frac{\mu\text{V}}{\sqrt{\text{Hz}}}\frac{0.1 \text{ mHz}}{f}$ for $f\ge 50\;\mu\text{Hz}$;
    \item {$\mathcal{R}_{\Delta_x,n}(\Delta t)\ll\mathcal{R}_{\Delta_x,dc}(\Delta t)$  at least for $\Delta t$ of the order of the days and months of  the  fluctuations of the potentials in Table~\ref{tab:deltax}. We capture this request by imposing that 
    \begin{equation*}
      \lim_{f\to 0}  S_{\Delta_x,n}^{1/2}(f)\ll\lim_{f\to 0}  S_{\Delta_x,dc}^{1/2}(f)
    \end{equation*}} with $S_{\Delta_x,dc}^{1/2}(f)$ the ASD corresponding to $\mathcal{R}_{\Delta_x,dc}(\Delta t)$.
\end{enumerate}

From these constraints, and using the joint posterior  for $\sigma_{\Delta_x,dc}$ and $\tau_{dc}$ to estimate $\lim_{f\to 0}  S_{\Delta_x,dc}^{1/2}(f)$, we get 
$2.5\times 10^4\text{ s}\leq \tau _n\leq 2.5\times 10^8 \text{ s}$ and $\sigma _{\Delta _x,n}\leq2\times 10^{-5} \frac{ \text{mV}}{\sqrt{s}}\sqrt{\tau_\text{n}}$.

Finally to summarize the above analysis, to make all functional forms explicit,  and to turn the results into  a model for  the coefficient $V_1^2(\Delta t)$, we write:
\begin{equation}
\label{eq:aone}
\begin{split}
&V_1^2(\Delta t)=\beta^2\frac{\alpha^4 }{\alpha ^4+1} \mathcal{R}_{\Delta_x}(\Delta t)\equiv\\
&\equiv V_{dc,1}^2 e^{-\frac{\lvert\Delta t\rvert}{\tau_{dc}}}\left(1+\frac{\lvert\Delta t\rvert}{\tau_{dc}}\right)+V_{n,1}^2e^{-\frac{\lvert\Delta t\rvert}{\tau_{n}}}
\end{split}
\end{equation}
with  parameter values  given in Table~\ref{tab:pv},

\begin{table}[h]
\caption{Parameter values for Eq.~\eqref{eq:aone}.}
\begin{equation*}
\begin{array}{|c|c|}
\hline
\text{Parameter} &\text{Value/Range} \\
\hline
V_{dc,1}&1.8_{-0.6}^{+1.2}\text{ mV}\\
\hline
\tau_{dc}&\left(3.7_{-1.3}^{+2.0}\right)\times 10^7\text{ s}\\
\hline
V_{n,1}&\le 3\times 10^{-6} \frac{\text{ mV}}{\sqrt{\text{s}}}\sqrt{\tau_n}\\
\hline
\tau_{n}&\in \lbrack2.5\times 10^4\text{ s},2.5\times 10^8 \text{ s}\rbrack\\
\hline
\end{array}
\end{equation*}
\label{tab:pv}
\end{table}

We finally note that with the parameters values in Table.~\ref{tab:pv}, $V_{dc,1}^2 e^{-\frac{\lvert\Delta t\rvert}{\tau_{dc}}}\left(1+\frac{\lvert\Delta t\rvert}{\tau_{dc}}\right)\gg V_{n,1}^2e^{-\frac{\lvert\Delta t\rvert}{\tau_{n}}}$ for all values of $\Delta t$.}

\section{\label{sect:proj}Projection of force noise  in LPF in the absence of charge bias}

We now use the results of the preceding section to estimate the force noise due to patch potentials in LPF. As the experimental results above only give information on $V_1^2(\Delta t)$, our estimate will require some modelling, and will be conditional to the specific model adopted. We will discuss a few possibilities, based on the available literature.

\subsection{Some model assumptions}

We will assume that, as for the case of $l=1$, the potential has both a quasi-static component, and an independent, much smaller   noise component, responsible for the patch potential fluctuation within the LPF frequency band. 

In addition we will assume that the slow drift behaviour observed in $V_1^2(\Delta t)$, is shared by the entire quasi-static component, while we leave open the frequency dependence of the single harmonic amplitudes of the noise component. We will discuss further the validity of such an assumption based on available evidence.

Under this assumption, the potential $V_K(\theta,\phi,t)$ has autocorrelation
\begin{equation}
\label{eq:twoauto}
\begin{split}
&\mathcal{R}_{V_K}(\omega,\Delta t)=f_2(\Delta t,\tau_{dc})\sum_{l=0}^\infty \frac{2 l+1}{4\pi}V_{dc,l}^2P_l(\cos(\omega))+\\
&+\sum_{l=0}^\infty \frac{2 l+1}{4\pi}V_{n,l}^2 (\Delta t)P_l(\cos(\omega))\equiv \\
&\equiv f_2(\Delta t,\tau_{dc})\mathcal{R}_{dc}(\omega)+\mathcal{R}_{n}(\omega,\Delta t)
\end{split}
\end{equation}
where  $V_{n,l}^2(\Delta t)\ll V_{dc,l}^2 f_2(\Delta t,\tau_{dc})$.

Within this model, to linear terms in $\mathcal{R}_{n}(\omega,\Delta t)$, the autocorrelation of the noisy part of the force becomes

\begin{equation}
\label{eq:fas}
\begin{split}
\mathcal{R}_F(\Delta t)=&\varepsilon_0^2 f_2(\Delta t,\tau_{dc})\sum _{l=1}^{\infty }K(l)\left(V_{dc,l}^2 V_{n,l+1}^2(\Delta t)+\right.\\&\left.+V_{dc,l+1}^2 V_{n,l}^2 (\Delta t)\right)
\end{split}
\end{equation}
plus a term proportional to the sum of terms containing $V_{dc,l}^2V_{dc,l+1}^2 f_2^2(\Delta t,\tau_{dc})$. One can check that under all hypotheses for the distribution of $V_{dc,l}^2$ discussed in the following, the contribution to the force PSD of the Fourier transform of this  term, which is $\propto 1/f^6$, is completely negligible.

% \subsection{Models for $V_{dc,l}$}
\subsection{Models for \texorpdfstring{$V_{dc,l}$}{Vdc,l} }

Let us now discuss possible models for $V_{dc,l}$ and then for $\mathcal{R}_{dc}(\omega)$, the angular autocorrelation of the quasi-static patch potentials. 

Not all isotropic autocorrelation functions that are viable on a flat surface are allowed on a sphere \cite{Huang2011}. Indeed, some of the coefficients $2\pi \int_0^\pi\mathcal{R}(\omega)P_l\left(\cos(\omega)\right)\sin(\omega)d\omega$, of the expansion of a generic one, $\mathcal{R}(\omega)$, in Legendre polynomials,  may turn to be negative and cannot then be taken as the  variance of the relative random harmonic amplitude.

Fortunately the exponential $\mathcal{R}(\omega)=\delta^2e^{-\omega/\omega_o}$, with $\delta^2$ the variance and  $\omega_o$ a positive angular scale  factor, the simplest model to describe correlation up to a given angular scale, is a viable one.  For such an autocorrelation, the coefficients of the expansion  are  independent of $l$ up to $\simeq 1/\omega_o$, and  decay rapidly after that.  If such a roll-off is well above $l\simeq 75$, where the coefficient $K(l)$   begins to display a significant attenuation then in practice $V_{dc,l} $ may be considered as a constant in the calculations of the force autocorrelation.

To try to understand if  a simple exponential model may describe the actual distribution of patches, and to get an idea of the length scale of the related exponentials, we now try a comparison with the observations available in the literature. 

We start  noting that there is  an analytic relation between the variance and the coefficient for $l=1$. If  $\mathcal{R}_{dc}(\omega)\simeq V_{K,rms}^2e^{-\omega/\omega_{dc}}$, we calculate that 
\begin{equation}
\label{eq:esdel}
V_{dc,1}^2=V_{K,rms}^2 2 \pi\frac{ \left(1-e^{-\frac{\pi }{\omega _{dc}}}\right) \omega _{dc}^2}{1+4 \omega _{dc}^2}
\end{equation}
Thus from our estimate of $V_{dc,1}$ we can get an estimate of $V_{K,rms}$ as a function of the coherence length $\omega_{dc}$. 

For $\omega_{dc}\ll 1/2$, and taking the most likely value for $V_{dc,1}$, $V_{dc,1}=1.8\text{ mV}$, the formula predicts $V_{K,rms}\simeq 23 \text{ mV} \times\left(1\text{ mm}/(R_{in}\omega_{dc})\right)$.

For the $\simeq 0.5\;\mu\text{m}$ kind of correlation length shown in the measurement of Ref.~\cite{patches2020}, this would predict $V_{K,rms}\simeq 40\text{ V}$ with a peak to peak variation, over the kind of samples considered there, in excess of 100 V. This is  orders  of magnitude larger than the observed fluctuations, that span at most 250 mV.

Thus the small scale patches detected in \cite{patches2020} are not related to the potential observed in LPF. In addition, if these have exponential autocorrelation, it is unlikely that the correlation length may be in the micrometre range.

A more detailed comparison can  be made to the results of Ref.~\cite{Robertson_2006}, that reports various scanning of 1 cm$\times$1 cm gold-coated samples. The scans were performed  at a much coarser resolution than those of Ref.~\cite{patches2020}, namely at steps of 0.5 mm in both directions, and with a 3  mm tip, which should give a spatial resolution of a fraction  of $\pm 3\text{ mm}/2$ \footnote{N. Robertson, private communication}. 

In particular Ref.~\cite{Robertson_2006} reports, in \emph{their} Fig.~3 the contour plots for  two Kelvin probe  scanning of two nominally equal  Au-coated samples taken at a pressure of $\simeq 10^{-7} \text{ mbar}$, not too different from that of LPF, a few months apart. The plots are quite different, one (marked February 2004) being dominated by some long range gradient-like pattern, the other (August 2004) having a more random appearance. 

\begin{figure*}[!htbp]
\includegraphics[width=.45 \textwidth]{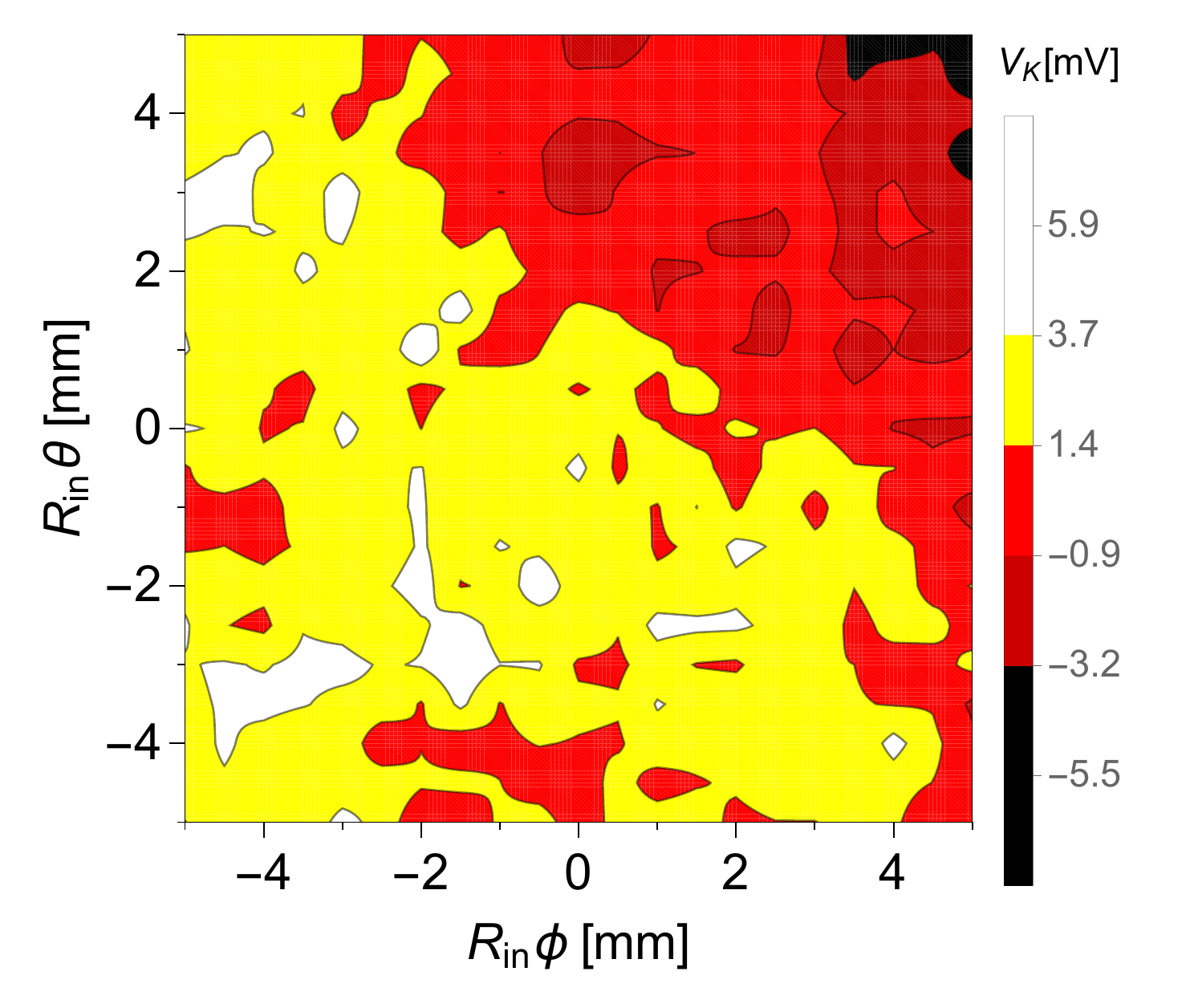}
\includegraphics[width=.45 \textwidth]{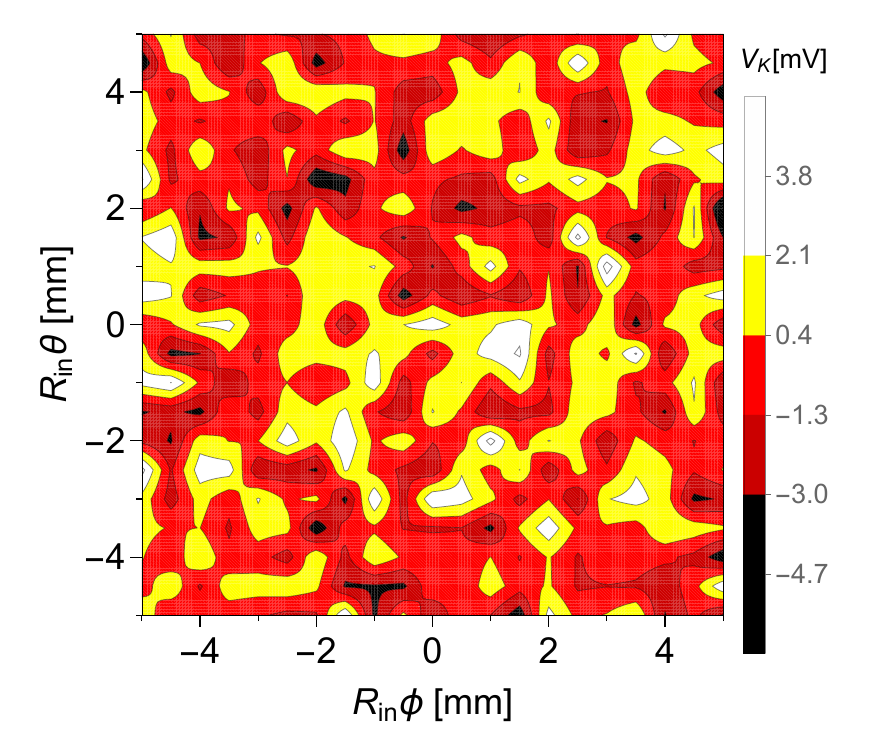}
\caption{Simulated patch potential distributions to be compared with  the Kelvin probe scans for the two Au-coated samples shown in  Ref.~\cite{Robertson_2006}. Left: simulated patch distribution with an exponential autocorrelation function $\mathcal{R}_{dc}=V_{K,rms}^2e^{-\omega/\omega_o}$, with $R_{in}\omega_o=10\text{ mm}$, and $V_{K,rms}$ adjusted to give $V_{dc,1}=1.7$\,mV. A random number from a Gaussian distribution with $\sigma_V=1\text{ mV}$ has been superimposed to each sample of the simulated scan. This plot is intended to mimic the upper left panel of Fig.~3 of  Ref.~\cite{Robertson_2006}. Right: contour plot for a matrix of random Gaussian numbers with $\sigma_V=1.7\text{ mV}$. This plot is intended to mimic the middle left panel of Fig.3 of  Ref.~\cite{Robertson_2006}. Both  plots show a projection of the domain of the TM surface with $R_{in}\theta \in \lbrack R_{in} \pi/2-5 \text{ mm},R_{in} \pi/2+5 \text{ mm}\rbrack$ and $R_{in}\phi \in \lbrack -5 \text{ mm},+5 \text{ mm}\rbrack$. }
\label{fig:patchcw}
\end{figure*}

We have qualitatively reproduced both scans, with the results shown in Fig.~\ref{fig:patchcw} and a brief description of the simulation input here below.  We stress that the simulation involves also the algorithm to produce the contour plot from the data, the results of which may affect the appearance of the plot. Thus the comparison is just merely qualitative.

Nevertheless, the August 2004 scan is reasonably well reproduced by a matrix of $20\times$20 independent  random numbers extracted by a Gaussian distribution with $\sigma_V=1.7$\,mV. Such would be the case both if the scan is limited by the noise of the technique, or if the effective resolution is better than the scanning step of 0.5 mm. In the second case, the scan would result from the averaging of a distribution with a correlation length $\ll0.5$\,mm, for which $V_{dc,1}\simeq 70\,\mu$V.

The February 2004 scan is reproduced instead with a distribution of patches with a correlation length $R_{in}\omega_o=10$ mm, and $V_{dc,1}=1.7$ mV, like in LPF,  superimposed to a random field, like in the previous case, but this time with   $\sigma_V\simeq 1 $ mV. 

The implication of the coincidence in the value of $V_{dc,1}$ in this simulation and that in LPF should not be overstated but still is an indication of a similar
range of phenomena. The lower value of $\sigma_V$, relative to the February scan, may be an indication that only part of the distribution in the latter might be due to instrumental uncertainty.

We want also to point out that Ref.~\cite{Robertson_2006} also observed  a voltage bias of the entire cm-sized sample in the hundreds of mV scale, in addition to those shown in the pictures.
This bias was observed to  change randomly, again by hundred mV, for the same sample, both from one installation into the vacuum chamber to the other, but also in different days but within the same installation.

These fluctuations are not included in the distribution we used for the simulations in Fig.~\ref{fig:patchcw} and, if due to patch potential and not to some other instrumental artefact, must be included by hand. 

As already mentioned, this may be a sign that samples get almost uniformly coated by contaminants, with small relative density/composition variations that account for the fluctuations of Fig.~\ref{fig:patchcw}.

A series of recent studies with torsion pendulums also provide some valuable information. A linear scanning of a gold-plated torsion pendulum planar blob, in steps of 0.125 mm, and with a probe 5 mm wide, showed mV variation on a $\simeq 5\text{ mm}$ scan \cite{Wuhan}. A crude periodogram analysis of the reported data gives indeed a $1/\nu^2$ behaviour, with $\nu$ the spatial frequency. This is what   one would expect from an exponential autocorrelation, if the coherence length is longer than $\simeq 5 \text{ mm}/(2 \pi)\simeq 0.8 \text{ mm}$, the inverse of the spectral resolution. 

In addition, a recent measurement  of the force gradient between a gold-coated plate suspended as a torsion pendulum element and a conducting membrane of cm dimensions \cite{Wuhan23}, points to a patch scale size $\simeq 0.5 \text { mm}$.

Finally, in discussing the available evidence on the distribution of $V_{dc,l}$, it is worth mentioning the observations of the patch induced torques on the gyroscopes of  the Gravity Probe B (GP-B) mission \cite{Buchman_2011}. Such observations, that refer though to  Nb-coated spheres, are compatible with $V_{dc,l}\propto l^{-n}$ with $n > 0$.  

Our tentative conclusion from this overview is that $\mathcal{R}_{dc}(\omega)$ in systems likely dominated by contamination,  may be compatible with an exponential shape, but the coherence length must be significantly larger than micrometres. The absolute scale of the phenomenon seems to put $V_{K,rms}$ in the 1-10 mV range, probably depending on the nature and the extent of the contaminants.

Thus, one can obtain a cautious upper limit for the force autocorrelation, by assuming a constant value for $V_{dc,l}^2$, $V_{dc,l}^2=V_{dc,1}^2$, instead of a decaying one because of the cut-off in the exponential or because of some power law behaviour. 
% \subsection{Models for $V_{n,l}^2 (\Delta t)$}
\subsection{Models for \texorpdfstring{$V_{n,l}^2 (\Delta t)$}{Vnl2(Dt)} }

Let us now review the available evidence on $\mathcal{R}_{n}(\omega,\Delta t)$ or, more in general, on the possible in-band fluctuations of the higher harmonics of the patch distribution. 

Experiments with torsion pendulums \cite{prlelectro,Pollack,Wuhan}, that suffer from the same limitation of resulting from mm-to cm-wide surface average measurements as for LPF, seem to give slightly conflicting information.

In particular Ref.~\cite{prlelectro}, a measurement of the torsional analogue of $\Delta_x$ on a LISA-like TM, did not detect any significant patch noise down to 0.1 mHz, and could only put an upper limit for its ASD, an upper limit  completely superseded by that from   LPF discussed above.

Ref.~\cite{Pollack},  using a plate shaped test-mass close (0.1-1\,mm) to a metal plate, did instead detect some noise, with a `surprisingly white' (in the words of the authors) ASD down to 0.1 mHz and $1/f$ tail in the ASD below that. This noise seems to exceed the upper limit from LPF discussed above, however the two experiments are different enough that a direct comparison might be misleading.

Ref.~\cite{Wuhan} reports a noise ASD of the potential of their 5\,mm probe of about $0.5 \text{ mV}/\sqrt{\text{Hz}}$, which seems definitely higher than the upper limit in \cite{prlelectro}. However the authors only give the ASD of the overall electrical potential and do not discuss the role of other possible sources of electrical noise.

An attempt of Ref.~\cite{Robertson_2006} to measure time fluctuations of the patch potentials turned out to be limited by the electronic noise of the Kelvin probe.

Some relevant information on electric field fluctuations nearby metal surfaces is available from studies on anomalous heating of ions in ion traps \cite{IonTraps}. Patch field fluctuations are considered there as one of the possible sources of electric field fluctuations responsible for the heating. 

Among all possible mechanisms that may make the patch potential to fluctuate,  diffusion limited fluctuation of the  density { of the contaminant adsorbate}, which is expected to exhibit a linear relation to electrostatic potential, is supported by some evidence, mostly the fact that it predicts  $1/f^{3/2}$ behaviour of the PSD of the electric field observed in some experiments \cite{IonTraps}. Though those experiments refer to the MHz frequency range, there is no obvious reason why such a spectrum should not extend down to low frequencies.

The model calculate the density autocorrelation on the surface $S$ of a planar electrode  as: 
\begin{equation}
\label{eq:diffprop}
\begin{split}
&\langle \delta n(\vec{r}_1, t+\Delta t)\delta n(\vec{r_2},t)\rangle=\\
&=\int_S G(\vec{r}_1, t+\Delta t, \vec{r}_3,t)\langle \delta n(\vec{r}_2, t)\delta n(\vec{r_3},t)\rangle d\vec{r}_3,
\end{split}
\end{equation}
with 
\begin{equation}
G(\vec{r}_1, t+\Delta t, \vec{r}_3,t)=e^{-\lvert \vec{r}_1- \vec{r}_3\rvert^2/(\tilde{D}\lvert \Delta t\rvert)}/ (\tilde{D} \lvert\Delta t \rvert)
\end{equation}
the Green function for diffusion on a plane, with diffusion coefficient $\tilde{D}$.

The analogous Green function for diffusion on a sphere of radius $R_{in}$ can be calculated to be  \cite{diffsphere,Ledesma}
\begin{equation}
\begin{split}
&G(\theta_1,\phi_1, t+\Delta t, \theta_3,\phi_3,t)=\\
&=\frac{1}{R_{in}^2}\sum_{l=0}^\infty \frac{2 l+1}{4 \pi}e^{-l(l+1)\lvert \Delta t\rvert/\tau_d}P_l(\cos(\omega))
\end{split}
\end{equation}
where, as usual, $\omega$ is the angular distance between the two points $(\theta_1,\phi_1)$ and $(\theta_3,\phi_3)$, and where $\tau_d=R_{in}^2/\tilde{D}$. 

Taking $V_K(\theta_1,\phi_1,t)\propto \delta n (\theta_1,\phi_1,t)$, then  $\langle \delta n(\theta_2,\phi_2, t)\delta n(\theta_3,\phi_3,t)\rangle\propto \mathcal{R}_n(\cos(\omega),0)$. By using the expansion of Legendre polynomial into spherical harmonics, the integral in Eq.~\eqref{eq:diffprop}, translated into our spherical geometry, can be performed explicitly, giving:
\begin{equation}
\label{eq:acdif}
\begin{split}
&\mathcal{R}_{n}(\omega,\Delta t)=\sum_{l=0}^\infty \frac{2 l+1}{4 \pi}V_{n,l}^2(0)e^{-l(l+1)\lvert \Delta t\rvert/\tau_d}P_l(\cos(\omega))
\end{split}
\end{equation}

We note that the model gives a well defined time scale $\tau_d$, proportional to the time for diffusion around the entire sphere. This is a remarkable difference with the infinite plane model where the analogous time scale becomes infinite.

Still on $\tau_d$, we also note that it takes two different values for the inner and the outer sphere, as the radii are different. The variation is about 20\% which, for the purpose of the current work,  is negligible compared to the relative uncertainties on all time parameters like $\tau_{dc}$. In the following calculation we will thus neglect the effect of this small inaccuracy.

By using Eq.~\eqref{eq:acdif} into  Eq.~\eqref{eq:fas}, Fourier transforming  and taking the square root, we can get the ASD of the predicted force noise as:
\begin{equation}
\label{eq:fasddummy}
\begin{split}
&S_F^{1/2}(f)\le\varepsilon_0V_{dc,1}V_{n,1}\times\\
&\times\left( \sum_{l=1}^\infty K(l) \left( \left(\frac{V_{dc,l+1}V_{n,l}}{V_{dc,1}V_{n,1}}\right)^2 g(l,f,\tau_{dc},\tau_d)\right.\right.+\\
&\left.\left.+  \left(\frac{V_{dc,l}V_{n,l+1}}{V_{dc,1}V_{n,1}}\right)^2g(l+1,f,\tau_{dc},\tau_d)\right)\right)^{1/2}
\end{split}
\end{equation} 
where $V_{n,l}^2\equiv V_{n,l}^2(0)$, $g(l,f,\tau_{dc})$ is the Fourier transform of $f_2(\Delta t,\tau_{dc})\times e^{-l(l+1)\lvert \Delta t\rvert/\tau_d}$, and where, finally,  the coefficients of $g(l,f,\tau_{dc})$ and $g(l+1,f,\tau_{dc})$  have been normalised to those for $l=1$ to make the sum independent of any common scale factor.

To proceed further we need the  dependency of $V_{n,l}^2$ on $l$, that is the shape of   $\mathcal{R}_{n}(\omega,0)$. Again the simple exponential decay seems general enough for the purpose of the present discussion. In addition one would expect that the density of diffusing adatoms should be proportional to the density of adsorbate, so that   $\mathcal{R}_{n}(\omega,0)\propto \mathcal{R}_{dc}(\omega) $. 

As said, with such an exponential distribution, the assumption that maximises the sum and then the ASD,  is that of a very short coherence length, that makes  $V_{n,l}/V_{n,1}=V_{dc,l}/V_{dc,1}=1$ for all values of $l$ well beyond the cutoff of $K(l)$. 

With this assumption in hand, we only need an estimate of  $V_{n,1}^2$ to calculate the ASD. Consider that the model in Eq.~\eqref{eq:acdif} predicts  that $V_{n,1}^2(\Delta t)=V_{n,1}^2 e^{-3 \lvert \Delta t\rvert/\tau_d}$ and that $S_{V_{n,1}}(f)\simeq 4/((2 \pi f)^2 \tau_d /3)$. 

Taking the maximum bound for such a spectral density,  $\tau_d \simeq 3 \tau_{dc}$ and $ V_{n,1}^2=( \tau_{dc} S_o (2 \pi \, 0.1 \text{ mHz})^2)/4$, the ASD in Eq.~\eqref{eq:fasddummy} can be calculated numerically using  values for $\tau_{dc}$ and $V_{dc,1}$ extracted from the posterior behind Table~\ref{tab:pv}. 

\begin{figure}[!htbp]
\includegraphics[width=1 \columnwidth]{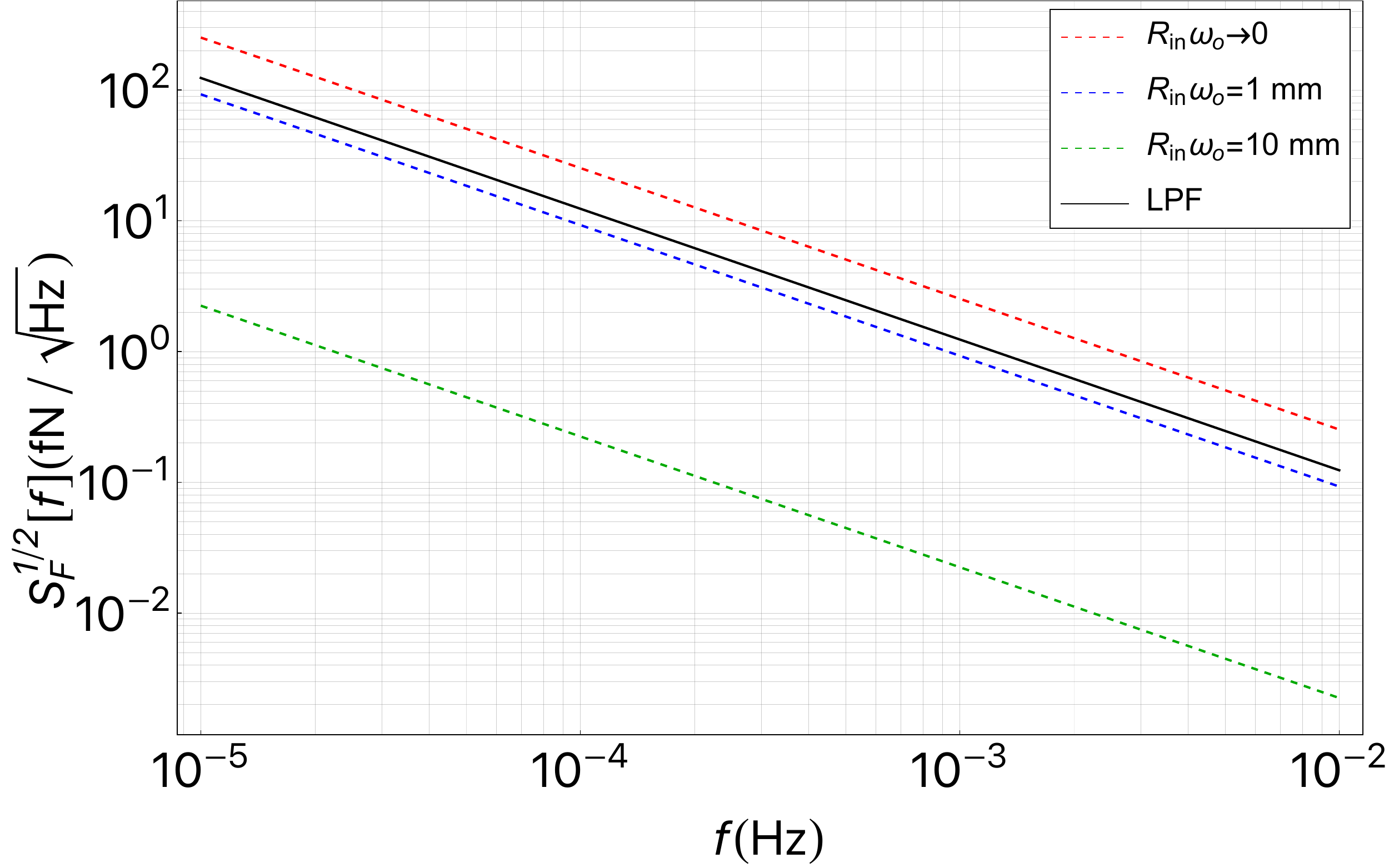}
\caption{Projected force noise ASD for a single test-mass from the diffusion model Eq.~\eqref{eq:fasddummy}, with $\tau_d=3\tau_{dc}$, $V_{n,1}$ from the upper bound for the LPF measured noise, and using the posterior for  $\tau_{dc}$ and $V_{dc,1}$   in Table~\ref{tab:pv}. Red dashed line: the case for  $V_{dc,l}/V_{dc,1}=V_{n,l}/V_{n,1}=1$. Blue dashed line: the case for both $V_{dc,l}/V_{dc,1}$ and $V_{n,l}/V_{n,1}$ calculated from an exponential correlation with a coherence length of 1 mm. Green dashed line: the case for both $V_{dc,l}/V_{dc,1}$ and $V_{n,l}/V_{n,1}$ calculated from an exponential correlation with a coherence length of 10 mm. Black line: best fit to $1/f$ function, of the ASD { of the  force noise in excess of the Brownian noise}, expressed as single test-mass disturbance, measured during the February 2017 noise run of LPF \cite{Ultimate}. {For reference, the value of the Brownian noise, not shown in the figure, is $\simeq 2 \text{ fN}/\sqrt{\text{Hz}}$.}  }
\label{fig:diff}
\end{figure}

We report in Fig.~\ref{fig:diff} the $+ 2\sigma$ upper bound to the force noise for such a calculation. For comparison the figure also includes the excess force noise ASD measured during the February 2017 noise run of LPF \cite{Ultimate}.

The picture clearly shows that the model predicts indeed a noise ASD with $1/f$ frequency dependence. However, the $+2\sigma$ upper limit predicted by this very conservative limit exceeds the total observed excess noise by almost a factor 2 in the limit of vanishing coherence length. 

We note that this conservative upper limit would be reduced considerably, even for small coherence length, if the observed in-band noise is indeed explained just by electronics noise even at the lowest frequency.

What also significantly reduces the value of the predicted noise is dropping the very conservative assumption of an infinitely short coherence length of the geometrical patch distribution. To illustrate the effect we also report in Fig.~\ref{fig:diff} the $+ 2\sigma$ upper limit  calculated under the hypothesis that $V_{n,l}\propto V_{dc,l}$ and that $\mathcal{R}_{dc}(\omega)\propto e^{-\omega/\omega_o}$, for both  $R_{in}\omega_o=10 \text{ mm}$, and $R_{in}\omega_o=1 \text{ mm}$, distributions  close to  our simulations  of Fig.~\ref{fig:patchcw}. The picture shows that the noise decreases non-linearly with increasing coherence length, becoming negligible compared to the LPF noise excess for $R_{in}\omega_o\simeq10 \text{ mm}$.

The assumption $\tau_d=3 \tau_{dc}$, may seem a bit unnatural, giving to the noisy part of the potential a similar time evolution as the quasi-dc one. However the posterior for the diffusion coefficient one gets from such a hypothesis has a $\pm 1\sigma$ range of $\tilde{D}=\left(1.4_{-0.7}^{+1.5}\right)\times 10^{-11} \text{ m}^2\text{s}^{-1}$. This is well within the range $10^{-15}-10^{-9}\text{ m}^2\text{s}^{-1}$ quoted by \cite{IonTraps}.  

One should also mention that, in the case of very short coherence length $V_{dc,l}/V_{dc,1}=V_{n,l}/V_{n,1}=1$, the effect of a shortened $\tau_d$ is that of bending the $1/f$ line of the predicted ASD, into a plateau at the lowest frequency. For instance, already   $\tau_d=0.3 \tau_{dc}$ would make the bending visible at around 0.1 mHz, without affecting the value of the $1/f$ tail at higher frequencies.

On the contrary,  the cutoff in $V_{dc,l}$ introduced by the long coherence length, makes also the $1/f$ behaviour less sensitive to $\tau_d$, so that even taking $\tau_d=0.03 \tau_{dc}$, does not affect significantly the shape of the ASD if $R_{in}\omega_o=10 \text{ mm}$. Note that $\tau_d=0.03 \tau_{dc}$ would give  $D=\left(1.4_{-0.7}^{+1.5}\right)\times 10^{-9} \text{ m}^2\text{s}^{-1}$, still within the range mentioned above.

\section{\label{sect:conc} Conclusions}
The concentric sphere model has proven to be a useful tool to analyse the effect of the fluctuation of patch potentials in the actual topology of a finite-size, three-dimensional test mass enclosed in a hollow housing. 

Nevertheless, its application to the current available information on the space distribution, and the time evolution of these potentials, from LPF and other relevant experiments, does not allow us to rule out that they may have contributed to the force noise detected in LPF, in excess of the Brownian plateau below $\simeq 1$ mHz. 

We have actually sketched a physically reasonable scenario of { contaminant} adsorbate patches with density fluctuations correlated over a length of millimetres, accompanied by diffusion of adatoms on their surface, which may have caused a noise with the right frequency dependence and the right order of magnitude, to explain all or part of the unexplained excess.

Though such scenario remains unproven, its possibility would suggest a few precautions to be followed in the development of LISA. 

First,  torsion pendulum experiments with LISA-like TM have achieved  sensitivities  \cite{prlelectro,Giuliana}  that may allow a direct detection of, or a significant upper limit to the noise we are discussing here. 

Such an experiment is under analysis, but would require reducing the gaps around the TM to about or less than
1 mm, considering the scaling with gap in Fig.~\ref{fig:kappa},  and would likely detect torque noise rather than force. Such a direct measurement would supersede much of the analysis presented here and provide a stronger experimental anchor to keep this potential noise source under control. 

In the absence of a direct measurement, a measure of precaution is to investigate the nature and the extent of the adsorbate that may have been present on the surface of LPF TM and EH during its operations.  The objective is that  of keeping TM contamination in LISA close to,  or better than that in LPF. 

This requires a campaign of surface characterisation on samples that have undergone a similar preparation history of that of LPF test-masses. A systematic experimental study, with the  Kelvin probe technique, of the quasi-static distribution of patch potentials, would be an important part of such a characterisation campaign.

It is reasonable to assume that if no new contaminants are introduced in LISA, that had not been present in LPF, and if the amount of contamination can be kept below that of LPF, the noise performance of LPF, that fulfils LISA requirements, may be confidently be reproduced.

\section*{Data availability statement}
All data that support the findings of this study are included within the article (and any supplementary information files).

\section*{Acknowledgements}
We thank Norna Robertson and Annika Lang for their patience and their very valuable feedback to our questions. We thank Vittorio Chiavegato for providing the posterior of the $1/f^2$ component of $\Delta_x$ noise.

This work has been made possible by the LISA Pathfinder mission, which is part of the space-science program of the European Space Agency, and has been supported by Istituto Nazionale di Fisica Nucleare (INFN) and Agenzia Spaziale Italiana (ASI), Project No. 2017-29-H.1-2020 ``Attivit\`a per la fase A della missione LISA''. % Italy

% \newpage
\appendix
{
% \section{\label{app:bayes} Bayesian fit to $\Delta_x(t)$ data}
\section{\label{app:bayes} Bayesian fit to \texorpdfstring{$\Delta_x(t)$}{Dx} data}

The physical model behind the fit assumes that $\Delta_{x,1}(t)$ and $\Delta_{x,2}(t)$ are two independent zero-mean Gaussian stochastic processes, both with the same auto-correlation $R_{\Delta_x}(\Delta t)\equiv\langle \Delta_x(t)\Delta_x(t+\Delta t)\rangle$. 

In addition, we assume that each entry $\Delta_{x,l}(t_i)$  in Table~\ref{tab:deltax} is the sum of a sample from one of these processes, and of one from an independent zero-mean Gaussian variable, representing the measurement error,  with a standard deviation equal to the error associated with $\Delta_{x,l}(t_i)$ in the table.

These hypotheses, and a model for $R_{\Delta_x}(\Delta t)$, suffice to write the theoretical joint covariance matrix $\mathbf{\Sigma}$ of the samples in Table~\ref{tab:deltax}. Since all variables are Gaussian and zero-mean, knowing $\mathbf{\Sigma}$ determines the entire joint probability density of the samples. According to the standard Bayesian approach, this probability density can then be used to construct the posterior likelihood of the observed data.

Given the independence of $\Delta_{x,1}(t)$ and $\Delta_{x,2}(t)$, $\mathbf{\Sigma}$ is a block matrix with a block $\mathbf{\Sigma}_1$ for the TM1 data and a block $\mathbf{\Sigma}_2$ for the TM2 data.  We now discuss how either of these blocks is composed.

We have considered three models for $R_{\Delta_x}(\Delta t)$, each identified by the negative exponent in the model power law dependence of the ASD of $\Delta_x$ on $f$ (1/2, 1 and 2):
\begin{equation}
\begin{split}
    \mathcal{R}_{\Delta_x,1/2}(\Delta t)&=\sigma_{\Delta_x}^2\frac{\Gamma\left(0,\frac{\lvert\Delta t\rvert}{\tau_\text{max}}\right)-\Gamma\left(0,\frac{\lvert\Delta t\rvert}{\tau_\text{min}}\right)}{\log(\frac{\tau_\text{max}}{\tau_\text{min}})}\equiv\\
    &\equiv\sigma_{\Delta_x}^2 f_{1/2}(\Delta t),
    \label{eq:effe12}
    \end{split}
\end{equation}
\begin{equation}
\begin{split}
    &\mathcal{R}_{\Delta_x,1}(\Delta t)=\sigma_{\Delta_x}^2 e^{-\lvert \Delta t\rvert/\tau}\equiv\sigma_{\Delta_x}^2 f_1(\Delta t,\tau),
    \end{split}
    \label{eq:effe1}
\end{equation}
and
\begin{equation}
\begin{split}
    \mathcal{R}_{\Delta_x,2}(\Delta t)&=\sigma_{\Delta_x}^2 e^{-\lvert \Delta t\rvert/\tau}\left(1+\lvert \Delta t\rvert/\tau\right)\equiv\\&\equiv \sigma_{\Delta_x}^2 f_2(\Delta t,\tau).
    \end{split}
    \label{eq:effe2}
\end{equation}
that correspond respectively to the following ASDs \footnote{We use the one-sided ASD, defined as the square root of twice the  Fourier transform of the autocorrelation.}
\begin{equation}
    \begin{split}
       &S_{\Delta_x,1/2}^{1/2}(f)=\sigma_{\Delta_x}\sqrt{\frac{2}{\pi }}\times\\
       &\times \sqrt{\frac{\tan ^{-1}\left(2 \pi  f \tau _{\max }\right)-\tan ^{-1}\left(2 \pi  f \tau _{\min }\right)}{f \log \left(\frac{\tau _{\max }}{\tau _{\min }}\right)}},
       \label{eq:effe12a}
    \end{split}
\end{equation}
\begin{equation}
    S_{\Delta_x,1}^{1/2}(f)=\sigma_{\Delta_x}\sqrt{\frac{4\tau} {1+\left(2 \pi f \tau\right)^2}},
    \label{eq:effe1a}
\end{equation}
and
\begin{equation}
    S_{\Delta_x,2}^{1/2}(f)=\sigma_{\Delta_x}\sqrt{\frac{8\tau} {\left(1+\left(2 \pi f \tau\right)^2\right)^2}}.
    \label{eq:effe2a}
\end{equation}

In Eqs.~\eqref{eq:effe12} to~\eqref{eq:effe2a}, $\Gamma(0,z)$ is the incomplete gamma function of 0 and $z$, while $\sigma_{\Delta_x}$, $\tau$, $\tau_\text{min}$ and $\tau_\text{max}$ are numerical parameters.

One can easily check that for $\tau_\text{min}f\ll1\ll\tau_\text{max}f$ and for $1\ll\tau f$
\begin{equation}
    \begin{split}
         &S_{\Delta_x,1/2}^{1/2}(f)\to\frac{\sigma_{\Delta_x}}{\sqrt{\log(\frac{\tau_\text{max}}{\tau_\text{min}})}}\times \frac{1}{f^{1/2}}\\
        &S_{\Delta_x,1}^{1/2}(f)\to \frac{\sigma_{\Delta_x}}{\pi \sqrt{\tau}}\times\frac{1}{f}\\
        &S_{\Delta_x}^{1/2}(f)\to\frac{\sigma_{\Delta_x}}{\sqrt{2} \pi^2 \tau^{3/2}}\times\frac{1}{f^2},
    \end{split}
    \label{eq:lim}
\end{equation}

For the fit with the first model we leave $\sigma_{\Delta_x}$ as a free fitting parameter while we take $\tau_\text{min}=1\text{ ms}$ and $\tau_\text{max}=1\text{ Gs}$, values that guarantee the condition $\tau_\text{min}f\ll1\ll\tau_\text{max}f$ at any frequency of interest.
For the fit with the other two models we leave both $\sigma_{\Delta_x}$ and $\tau$, as free fitting parameters.

Once the model has been chosen,  the resulting  $\mathbf{\Sigma}_l$ becomes a function of the fitting parameters.

For instance, in the case of the  model in Eq.~\eqref{eq:effe1}, $\mathbf{\Sigma}_l$ has elements
\begin{equation}
\begin{split}
    &\Sigma_{l,i,j}(\sigma_{\Delta_x}^2,\tau)=\langle \Delta_{x,l}(t_i)\Delta_{x,l}(t_j)\rangle=\\
    &=\sigma_{\Delta_x}^2 e^{-\lvert t_i-t_j\rvert/\tau}+\delta_{i,j}\sigma_{l,i}^2
\end{split}
\end{equation}
with  $\delta_{i,j}$  the Kronecker delta of $i$ and $j$, and $\sigma_{l,i}$  the measurement error on $\Delta_{x,l}(t_i)$.

The logarithm of the fitting likelihood $\Lambda$ can then be written, still for the case of the  model in Eq.~\eqref{eq:effe1},  as: 
\begin{equation}
    \begin{split}
 &\log(\Lambda)=-\frac{1}{2}\times\bigg(\log(\lvert \mathbf{\Sigma}_1(\sigma_{\Delta_x}^2,\tau)\rvert)+\log(\lvert \mathbf{\Sigma}_2(\sigma_{\Delta_x}^2,\tau)\rvert)+\\
 &+\sum_{i,j}\Sigma_{1,i,j}^{-1}(\sigma_{\Delta_x}^2,\tau)\Delta_{x,1}(t_i)\Delta_{x,1}(t_j)+\\
 &+\sum_{i,j}\Sigma_{2,i,j}^{-1}(\sigma_{\Delta_x}^2,\tau)\Delta_{x,2}(t_i)\Delta_{x,2}(t_j)\bigg)+\\
 &+\log(f(\sigma_{\Delta_x}^2,\tau))
    \end{split}
    \label{eq:like}
\end{equation}
where $f(\sigma_{\Delta_x}^2,\tau)$ is the  prior probability density for the two fitting parameters $\sigma_{\Delta_x}^2$ and $\tau$. 

Our choice for  $f(\sigma_{\Delta_x}^2,\tau)$ is to  consider $\sigma_{\Delta_x}^2$ and $\tau$ as independent, and their logarithms to be uniformly distributed. Thus, in practice, $\log(f(\sigma_{\Delta_x}^2,\tau))$ is a constant and we run our Monte Carlo Markov Chain search independently in the $\log(\sigma_{\Delta_x}^2)$ and in the $\log(\tau)$ domains.

For the other two models the procedure is the same except that for the model in Eq.~\eqref{eq:effe12} the search takes place  only in the $\log(\sigma_{\Delta_x}^2)$ domain.

The search produces a joint posterior distribution for the fitting parameters. For each sample of this distribution we calculate the corresponding ASD using Eq.~\eqref{eq:lim} and we build the posterior ditributions for the projected ASD that are summarized in Table~\ref{tab:projn}. In the table, the value in each entry is the median of the corresponding   posterior, while  errors span the 0.16-0.84 probability interval, i.e. the $\pm 1\sigma$ probability interval for a Gaussian distribution.

A summary of the   results of the fit in term of  parameter values is shown in Table~\ref{tab:fitp}.
\begin{table}[h]
    \centering
    \caption{Credible intervals for the fitting parameters for the three considered models. For each parameter, the value is the median of its marginal posterior, while  errors span the 0.16-0.84 probability interval, i.e. the $\pm 1\sigma$ probability interval for a Gaussian distribution.}
   $\begin{array}{|c|c|c|}
   \hline
       \text{ Model}&\sigma_{\Delta_x}\text{ (mV)} &\tau \text{ (s)} \\
       \hline
        \sigma_{\Delta_x}^2\times f_{1/2}(\Delta t) & 9.7_{-2.1}^{+3.2}&\text{-}\\
        \hline
        \sigma_{\Delta_x}^2\times f_1(\Delta t,\tau) & 17._{-6.}^{+17.}&\left(1.8_{-1.3}^{+6.}\right)\times 10^9\\
        \hline
        \sigma_{\Delta_x}^2\times f_2(\Delta t,\tau) & 14._{-4.}^{+9.}&\left(3.7_{-1.3}^{+2.1}\right)\times 10^7\\
        \hline
    \end{array}$
    \label{tab:fitp}
\end{table}

As usual with Bayesian fit, we have performed a posterior predictive test \cite{gelman2020bayesian} to assess its `goodness'. We generate an ensemble of random data predicted by our posterior: we pick a random sample of the posterior for the fitting parameter, calculate the corresponding $\Sigma$, and use it to generate simulated Gaussian data. $N$ repetitions of this procedure generate a random sample of data consistent with the posterior. 

On each of these samples we calculate the likelihood $\Lambda$  for the values of the parameters that maximize $\Lambda$ for the real data, and calculate for  which fraction $p$ of the sample we get $\Lambda \le \Lambda_\text{data}$, with $\Lambda_\text{data}$ being the value for the real data.

The test shows that all fits have comparable quality with  $p>0.45$ for all models.

}

\clearpage
\bibliography{generic.bib}

\begin{thebibliography}{30}%
\makeatletter
\providecommand \@ifxundefined [1]{%
 \@ifx{#1\undefined}
}%
\providecommand \@ifnum [1]{%
 \ifnum #1\expandafter \@firstoftwo
 \else \expandafter \@secondoftwo
 \fi
}%
\providecommand \@ifx [1]{%
 \ifx #1\expandafter \@firstoftwo
 \else \expandafter \@secondoftwo
 \fi
}%
\providecommand \natexlab [1]{#1}%
\providecommand \enquote  [1]{``#1''}%
\providecommand \bibnamefont  [1]{#1}%
\providecommand \bibfnamefont [1]{#1}%
\providecommand \citenamefont [1]{#1}%
\providecommand \href@noop [0]{\@secondoftwo}%
\providecommand \href [0]{\begingroup \@sanitize@url \@href}%
\providecommand \@href[1]{\@@startlink{#1}\@@href}%
\providecommand \@@href[1]{\endgroup#1\@@endlink}%
\providecommand \@sanitize@url [0]{\catcode `\\12\catcode `\$12\catcode `\&12\catcode `\#12\catcode `\^12\catcode `\_12\catcode `\%12\relax}%
\providecommand \@@startlink[1]{}%
\providecommand \@@endlink[0]{}%
\providecommand \url  [0]{\begingroup\@sanitize@url \@url }%
\providecommand \@url [1]{\endgroup\@href {#1}{\urlprefix }}%
\providecommand \urlprefix  [0]{URL }%
\providecommand \Eprint [0]{\href }%
\providecommand \doibase [0]{https://doi.org/}%
\providecommand \selectlanguage [0]{\@gobble}%
\providecommand \bibinfo  [0]{\@secondoftwo}%
\providecommand \bibfield  [0]{\@secondoftwo}%
\providecommand \translation [1]{[#1]}%
\providecommand \BibitemOpen [0]{}%
\providecommand \bibitemStop [0]{}%
\providecommand \bibitemNoStop [0]{.\EOS\space}%
\providecommand \EOS [0]{\spacefactor3000\relax}%
\providecommand \BibitemShut  [1]{\csname bibitem#1\endcsname}%
\let\auto@bib@innerbib\@empty
%</preamble>
\bibitem [{\citenamefont {Vitale}(2014)}]{Vitale}%
  \BibitemOpen
  \bibfield  {author} {\bibinfo {author} {\bibfnamefont {S.}~\bibnamefont {Vitale}},\ }\bibfield  {title} {\bibinfo {title} {{Space-borne gravitational wave observatories}},\ }\href {https://doi.org/10.1007/s10714-014-1730-2} {\bibfield  {journal} {\bibinfo  {journal} {General Relativity and Gravitation}\ }\textbf {\bibinfo {volume} {46}},\ \bibinfo {eid} {1730} (\bibinfo {year} {2014})}\BibitemShut {NoStop}%
\bibitem [{\citenamefont {Armano}\ \emph {et~al.}(2018)\citenamefont {Armano} \emph {et~al.}}]{Ultimate}%
  \BibitemOpen
  \bibfield  {author} {\bibinfo {author} {\bibfnamefont {M.}~\bibnamefont {Armano}} \emph {et~al.},\ }\bibfield  {title} {\bibinfo {title} {{Beyond the Required LISA Free-Fall Performance: New LISA Pathfinder Results down to $20\text{ }\text{ }\ensuremath{\mu}\mathrm{Hz}$}},\ }\href {https://doi.org/10.1103/PhysRevLett.120.061101} {\bibfield  {journal} {\bibinfo  {journal} {Phys. Rev. Lett.}\ }\textbf {\bibinfo {volume} {120}},\ \bibinfo {pages} {061101} (\bibinfo {year} {2018})}\BibitemShut {NoStop}%
\bibitem [{\citenamefont {Weber}\ \emph {et~al.}(2003)\citenamefont {Weber}, \citenamefont {Bortoluzzi}, \citenamefont {Cavalleri}, \citenamefont {Carbone}, \citenamefont {Da~Lio}, \citenamefont {Dolesi}, \citenamefont {Fontana}, \citenamefont {Hoyle}, \citenamefont {Hueller},\ and\ \citenamefont {Vitale}}]{Spie}%
  \BibitemOpen
  \bibfield  {author} {\bibinfo {author} {\bibfnamefont {W.~J.}\ \bibnamefont {Weber}}, \bibinfo {author} {\bibfnamefont {D.}~\bibnamefont {Bortoluzzi}}, \bibinfo {author} {\bibfnamefont {A.}~\bibnamefont {Cavalleri}}, \bibinfo {author} {\bibfnamefont {L.}~\bibnamefont {Carbone}}, \bibinfo {author} {\bibfnamefont {M.}~\bibnamefont {Da~Lio}}, \bibinfo {author} {\bibfnamefont {R.}~\bibnamefont {Dolesi}}, \bibinfo {author} {\bibfnamefont {G.}~\bibnamefont {Fontana}}, \bibinfo {author} {\bibfnamefont {C.~D.}\ \bibnamefont {Hoyle}}, \bibinfo {author} {\bibfnamefont {M.}~\bibnamefont {Hueller}},\ and\ \bibinfo {author} {\bibfnamefont {S.}~\bibnamefont {Vitale}},\ }\bibinfo {title} {{Position sensors for flight testing of LISA drag-free control}},\ in\ \href@noop {} {\emph {\bibinfo {booktitle} {Gravitational-Wave Detection}}},\ \bibinfo {series} {Proceedings of the Society of Photo-Optical Instrumentation Engineers (Spie)}, Vol.\ \bibinfo {volume} {4856},\ \bibinfo {editor} {edited by\ \bibinfo {editor}
  {\bibfnamefont {M.}~\bibnamefont {Cruise}}\ and\ \bibinfo {editor} {\bibfnamefont {P.}~\bibnamefont {Saulson}}}\ (\bibinfo {year} {2003})\ pp.\ \bibinfo {pages} {31--42}\BibitemShut {NoStop}%
\bibitem [{\citenamefont {Armano}\ \emph {et~al.}(2024{\natexlab{a}})\citenamefont {Armano} \emph {et~al.}}]{armano2024nanonewton}%
  \BibitemOpen
  \bibfield  {author} {\bibinfo {author} {\bibfnamefont {M.}~\bibnamefont {Armano}} \emph {et~al.} (\bibinfo {collaboration} {LISA Pathfinder Collaboration}),\ }\bibfield  {title} {\bibinfo {title} {{Nano-Newton electrostatic force actuators for femto-Newton-sensitive measurements: System performance test in the LISA Pathfinder mission}},\ }\href {https://doi.org/10.1103/PhysRevD.109.102009} {\bibfield  {journal} {\bibinfo  {journal} {Phys. Rev. D}\ }\textbf {\bibinfo {volume} {109}},\ \bibinfo {pages} {102009} (\bibinfo {year} {2024}{\natexlab{a}})}\BibitemShut {NoStop}%
\bibitem [{\citenamefont {Antonucci}\ \emph {et~al.}(2012)\citenamefont {Antonucci}, \citenamefont {Cavalleri}, \citenamefont {Dolesi}, \citenamefont {Hueller}, \citenamefont {Nicolodi}, \citenamefont {Tu}, \citenamefont {Vitale},\ and\ \citenamefont {Weber}}]{prlelectro}%
  \BibitemOpen
  \bibfield  {author} {\bibinfo {author} {\bibfnamefont {F.}~\bibnamefont {Antonucci}}, \bibinfo {author} {\bibfnamefont {A.}~\bibnamefont {Cavalleri}}, \bibinfo {author} {\bibfnamefont {R.}~\bibnamefont {Dolesi}}, \bibinfo {author} {\bibfnamefont {M.}~\bibnamefont {Hueller}}, \bibinfo {author} {\bibfnamefont {D.}~\bibnamefont {Nicolodi}}, \bibinfo {author} {\bibfnamefont {H.~B.}\ \bibnamefont {Tu}}, \bibinfo {author} {\bibfnamefont {S.}~\bibnamefont {Vitale}},\ and\ \bibinfo {author} {\bibfnamefont {W.~J.}\ \bibnamefont {Weber}},\ }\bibfield  {title} {\bibinfo {title} {{Interaction between Stray Electrostatic Fields and a Charged Free-Falling Test Mass}},\ }\href {https://doi.org/10.1103/PhysRevLett.108.181101} {\bibfield  {journal} {\bibinfo  {journal} {Phys. Rev. Lett.}\ }\textbf {\bibinfo {volume} {108}},\ \bibinfo {pages} {181101} (\bibinfo {year} {2012})}\BibitemShut {NoStop}%
\bibitem [{\citenamefont {Armano}\ \emph {et~al.}(2017)\citenamefont {Armano} \emph {et~al.}}]{Charge2017}%
  \BibitemOpen
  \bibfield  {author} {\bibinfo {author} {\bibfnamefont {M.}~\bibnamefont {Armano}} \emph {et~al.},\ }\bibfield  {title} {\bibinfo {title} {{Charge-Induced Force Noise on Free-Falling Test Masses: Results from LISA Pathfinder}},\ }\href {https://doi.org/10.1103/PhysRevLett.118.171101} {\bibfield  {journal} {\bibinfo  {journal} {Phys. Rev. Lett.}\ }\textbf {\bibinfo {volume} {118}},\ \bibinfo {pages} {171101} (\bibinfo {year} {2017})}\BibitemShut {NoStop}%
\bibitem [{\citenamefont {Antonucci}\ and\ \citenamefont {et~al}(2011)}]{Fromlab}%
  \BibitemOpen
  \bibfield  {author} {\bibinfo {author} {\bibfnamefont {F.}~\bibnamefont {Antonucci}}\ and\ \bibinfo {author} {\bibnamefont {et~al}},\ }\bibfield  {title} {\bibinfo {title} {{From laboratory experiments to LISA Pathfinder: achieving LISA geodesic motion}},\ }\href@noop {} {\bibfield  {journal} {\bibinfo  {journal} {Classical and Quantum Gravity}\ }\textbf {\bibinfo {volume} {28}} (\bibinfo {year} {2011})}\BibitemShut {NoStop}%
\bibitem [{\citenamefont {Speake}(1996)}]{Speake96}%
  \BibitemOpen
  \bibfield  {author} {\bibinfo {author} {\bibfnamefont {C.~C.}\ \bibnamefont {Speake}},\ }\bibfield  {title} {\bibinfo {title} {{Forces and force gradients due to patch fields and contact-potential differences}},\ }\href {https://doi.org/10.1088/0264-9381/13/11A/039} {\bibfield  {journal} {\bibinfo  {journal} {Classical and Quantum Gravity}\ }\textbf {\bibinfo {volume} {13}},\ \bibinfo {pages} {A291} (\bibinfo {year} {1996})}\BibitemShut {NoStop}%
\bibitem [{\citenamefont {Buchman}\ and\ \citenamefont {Turneaure}(2011)}]{Buchman_2011}%
  \BibitemOpen
  \bibfield  {author} {\bibinfo {author} {\bibfnamefont {S.}~\bibnamefont {Buchman}}\ and\ \bibinfo {author} {\bibfnamefont {J.~P.}\ \bibnamefont {Turneaure}},\ }\bibfield  {title} {\bibinfo {title} {{The effects of patch-potentials on the gravity probe B gyroscopes}},\ }\href {https://doi.org/10.1063/1.3608615} {\bibfield  {journal} {\bibinfo  {journal} {Review of Scientific Instruments}\ }\textbf {\bibinfo {volume} {82}},\ \bibinfo {pages} {074502} (\bibinfo {year} {2011})},\ \Eprint {https://arxiv.org/abs/https://pubs.aip.org/aip/rsi/article-pdf/doi/10.1063/1.3608615/16007578/074502\_1\_online.pdf} {https://pubs.aip.org/aip/rsi/article-pdf/doi/10.1063/1.3608615/16007578/074502\_1\_online.pdf} \BibitemShut {NoStop}%
\bibitem [{Note1()}]{Note1}%
  \BibitemOpen
  \bibinfo {note} {{The problem of GPB \cite {Buchman_2011} was the coupling of quasi-static patch potentials on the surface of their spherical gyroscopes, with the quasi-static patch potentials on the gyroscope housing. With slow modulations caused by spacecraft motion around the gyros, this coupling exerted torques and caused gyroscope spurious precession that limited the accuracy of the measurements. Patch potential fluctuations were not a limiting factor for the experiment.}}\BibitemShut {Stop}%
\bibitem [{\citenamefont {Pollack}\ \emph {et~al.}(2008)\citenamefont {Pollack}, \citenamefont {Schlamminger},\ and\ \citenamefont {Gundlach}}]{Pollack}%
  \BibitemOpen
  \bibfield  {author} {\bibinfo {author} {\bibfnamefont {S.~E.}\ \bibnamefont {Pollack}}, \bibinfo {author} {\bibfnamefont {S.}~\bibnamefont {Schlamminger}},\ and\ \bibinfo {author} {\bibfnamefont {J.~H.}\ \bibnamefont {Gundlach}},\ }\bibfield  {title} {\bibinfo {title} {{Temporal Extent of Surface Potentials between Closely Spaced Metals}},\ }\href {https://doi.org/10.1103/PhysRevLett.101.071101} {\bibfield  {journal} {\bibinfo  {journal} {Phys. Rev. Lett.}\ }\textbf {\bibinfo {volume} {101}},\ \bibinfo {pages} {071101} (\bibinfo {year} {2008})}\BibitemShut {NoStop}%
\bibitem [{\citenamefont {Jackson}(1975)}]{Jackson}%
  \BibitemOpen
  \bibfield  {author} {\bibinfo {author} {\bibfnamefont {J.~D.}\ \bibnamefont {Jackson}},\ }\href {https://cds.cern.ch/record/100964} {\emph {\bibinfo {title} {{Classical electrodynamics; 2nd ed.}}}}\ (\bibinfo  {publisher} {Wiley},\ \bibinfo {address} {New York, NY},\ \bibinfo {year} {1975})\BibitemShut {NoStop}%
\bibitem [{\citenamefont {Robertson}\ \emph {et~al.}(2006)\citenamefont {Robertson}, \citenamefont {Blackwood}, \citenamefont {Buchman}, \citenamefont {Byer}, \citenamefont {Camp}, \citenamefont {Gill}, \citenamefont {Hanson}, \citenamefont {Williams},\ and\ \citenamefont {Zhou}}]{Robertson_2006}%
  \BibitemOpen
  \bibfield  {author} {\bibinfo {author} {\bibfnamefont {N.~A.}\ \bibnamefont {Robertson}}, \bibinfo {author} {\bibfnamefont {J.~R.}\ \bibnamefont {Blackwood}}, \bibinfo {author} {\bibfnamefont {S.}~\bibnamefont {Buchman}}, \bibinfo {author} {\bibfnamefont {R.~L.}\ \bibnamefont {Byer}}, \bibinfo {author} {\bibfnamefont {J.}~\bibnamefont {Camp}}, \bibinfo {author} {\bibfnamefont {D.}~\bibnamefont {Gill}}, \bibinfo {author} {\bibfnamefont {J.}~\bibnamefont {Hanson}}, \bibinfo {author} {\bibfnamefont {S.}~\bibnamefont {Williams}},\ and\ \bibinfo {author} {\bibfnamefont {P.}~\bibnamefont {Zhou}},\ }\bibfield  {title} {\bibinfo {title} {{Kelvin probe measurements: investigations of the patch effect with applications to ST-7 and LISA}},\ }\href {https://doi.org/10.1088/0264-9381/23/7/026} {\bibfield  {journal} {\bibinfo  {journal} {Classical and Quantum Gravity}\ }\textbf {\bibinfo {volume} {23}},\ \bibinfo {pages} {2665} (\bibinfo {year} {2006})}\BibitemShut {NoStop}%
\bibitem [{\citenamefont {Lang}\ and\ \citenamefont {Schwab}(2015)}]{spherical}%
  \BibitemOpen
  \bibfield  {author} {\bibinfo {author} {\bibfnamefont {A.}~\bibnamefont {Lang}}\ and\ \bibinfo {author} {\bibfnamefont {C.}~\bibnamefont {Schwab}},\ }\bibfield  {title} {\bibinfo {title} {{Isotropic Gaussian random fields on the sphere: Regularity, fast simulation and stochastic partial differential equations}},\ }\href {https://doi.org/10.1214/14-AAP1067} {\bibfield  {journal} {\bibinfo  {journal} {The Annals of Applied Probability}\ }\textbf {\bibinfo {volume} {25}},\ \bibinfo {pages} {3047 } (\bibinfo {year} {2015})}\BibitemShut {NoStop}%
\bibitem [{\citenamefont {Armano}\ \emph {et~al.}(2016)\citenamefont {Armano} \emph {et~al.}}]{subfemtog}%
  \BibitemOpen
  \bibfield  {author} {\bibinfo {author} {\bibfnamefont {M.}~\bibnamefont {Armano}} \emph {et~al.},\ }\bibfield  {title} {\bibinfo {title} {{Sub-Femto-$g$ Free Fall for Space-Based Gravitational Wave Observatories: LISA Pathfinder Results}},\ }\href {https://doi.org/10.1103/PhysRevLett.116.231101} {\bibfield  {journal} {\bibinfo  {journal} {Phys. Rev. Lett.}\ }\textbf {\bibinfo {volume} {116}},\ \bibinfo {pages} {231101} (\bibinfo {year} {2016})}\BibitemShut {NoStop}%
\bibitem [{\citenamefont {Anza}\ and\ \citenamefont {et~al.}(2005)}]{LPF}%
  \BibitemOpen
  \bibfield  {author} {\bibinfo {author} {\bibfnamefont {S.}~\bibnamefont {Anza}}\ and\ \bibinfo {author} {\bibnamefont {et~al.}},\ }\bibfield  {title} {\bibinfo {title} {{The LTP experiment on the LISA Pathfinder mission}},\ }\href {https://doi.org/10.1088/0264-9381/22/10/001} {\bibfield  {journal} {\bibinfo  {journal} {Classical and Quantum Gravity}\ }\textbf {\bibinfo {volume} {22}},\ \bibinfo {pages} {S125} (\bibinfo {year} {2005})}\BibitemShut {NoStop}%
\bibitem [{\citenamefont {Armano}\ \emph {et~al.}(2024{\natexlab{b}})\citenamefont {Armano} \emph {et~al.}}]{armano2024}%
  \BibitemOpen
  \bibfield  {author} {\bibinfo {author} {\bibfnamefont {M.}~\bibnamefont {Armano}} \emph {et~al.} (\bibinfo {collaboration} {LISA Pathfinder Collaboration}),\ }\bibfield  {title} {\bibinfo {title} {{In-depth analysis of LISA Pathfinder performance results: Time evolution, noise projection, physical models, and implications for LISA}},\ }\href {https://doi.org/10.1103/PhysRevD.110.042004} {\bibfield  {journal} {\bibinfo  {journal} {Phys. Rev. D}\ }\textbf {\bibinfo {volume} {110}},\ \bibinfo {pages} {042004} (\bibinfo {year} {2024}{\natexlab{b}})}\BibitemShut {NoStop}%
\bibitem [{Note2()}]{Note2}%
  \BibitemOpen
  \bibinfo {note} {By `in-band' we mean here within the 0.02-1 mHz frequency range}\BibitemShut {NoStop}%
\bibitem [{Note3()}]{Note3}%
  \BibitemOpen
  \bibinfo {note} {This figure is taken from the posterior distribution behind the fit to the in-band noise of figure 9 of Ref.~\cite {armano2024nanonewton}.}\BibitemShut {Stop}%
\bibitem [{\citenamefont {Huang}\ \emph {et~al.}(2011)\citenamefont {Huang}, \citenamefont {Zhang},\ and\ \citenamefont {Robeson}}]{Huang2011}%
  \BibitemOpen
  \bibfield  {author} {\bibinfo {author} {\bibfnamefont {C.}~\bibnamefont {Huang}}, \bibinfo {author} {\bibfnamefont {H.}~\bibnamefont {Zhang}},\ and\ \bibinfo {author} {\bibfnamefont {S.~M.}\ \bibnamefont {Robeson}},\ }\bibfield  {title} {\bibinfo {title} {{On the Validity of Commonly Used Covariance and Variogram Functions on the Sphere}},\ }\href {https://doi.org/10.1007/s11004-011-9344-7} {\bibfield  {journal} {\bibinfo  {journal} {Mathematical Geosciences}\ }\textbf {\bibinfo {volume} {43}},\ \bibinfo {pages} {721} (\bibinfo {year} {2011})}\BibitemShut {NoStop}%
\bibitem [{\citenamefont {Garrett}\ \emph {et~al.}(2020)\citenamefont {Garrett}, \citenamefont {Kim},\ and\ \citenamefont {Munday}}]{patches2020}%
  \BibitemOpen
  \bibfield  {author} {\bibinfo {author} {\bibfnamefont {J.~L.}\ \bibnamefont {Garrett}}, \bibinfo {author} {\bibfnamefont {J.}~\bibnamefont {Kim}},\ and\ \bibinfo {author} {\bibfnamefont {J.~N.}\ \bibnamefont {Munday}},\ }\bibfield  {title} {\bibinfo {title} {{Measuring the effect of electrostatic patch potentials in Casimir force experiments}},\ }\href {https://doi.org/10.1103/PhysRevResearch.2.023355} {\bibfield  {journal} {\bibinfo  {journal} {Phys. Rev. Res.}\ }\textbf {\bibinfo {volume} {2}},\ \bibinfo {pages} {023355} (\bibinfo {year} {2020})}\BibitemShut {NoStop}%
\bibitem [{Note4()}]{Note4}%
  \BibitemOpen
  \bibinfo {note} {N. Robertson, private communication}\BibitemShut {NoStop}%
\bibitem [{\citenamefont {Yin}\ \emph {et~al.}(2014)\citenamefont {Yin}, \citenamefont {Bai}, \citenamefont {Hu}, \citenamefont {Liu}, \citenamefont {Luo}, \citenamefont {Tan}, \citenamefont {Yeh},\ and\ \citenamefont {Zhou}}]{Wuhan}%
  \BibitemOpen
  \bibfield  {author} {\bibinfo {author} {\bibfnamefont {H.}~\bibnamefont {Yin}}, \bibinfo {author} {\bibfnamefont {Y.-Z.}\ \bibnamefont {Bai}}, \bibinfo {author} {\bibfnamefont {M.}~\bibnamefont {Hu}}, \bibinfo {author} {\bibfnamefont {L.}~\bibnamefont {Liu}}, \bibinfo {author} {\bibfnamefont {J.}~\bibnamefont {Luo}}, \bibinfo {author} {\bibfnamefont {D.-Y.}\ \bibnamefont {Tan}}, \bibinfo {author} {\bibfnamefont {H.-C.}\ \bibnamefont {Yeh}},\ and\ \bibinfo {author} {\bibfnamefont {Z.-B.}\ \bibnamefont {Zhou}},\ }\bibfield  {title} {\bibinfo {title} {{Measurements of temporal and spatial variation of surface potential using a torsion pendulum and a scanning conducting probe}},\ }\href {https://doi.org/10.1103/PhysRevD.90.122001} {\bibfield  {journal} {\bibinfo  {journal} {Phys. Rev. D}\ }\textbf {\bibinfo {volume} {90}},\ \bibinfo {pages} {122001} (\bibinfo {year} {2014})}\BibitemShut {NoStop}%
\bibitem [{\citenamefont {Ke}\ \emph {et~al.}(2023)\citenamefont {Ke}, \citenamefont {Dong}, \citenamefont {Huang}, \citenamefont {Tan}, \citenamefont {Tan}, \citenamefont {Yang}, \citenamefont {Shao},\ and\ \citenamefont {Luo}}]{Wuhan23}%
  \BibitemOpen
  \bibfield  {author} {\bibinfo {author} {\bibfnamefont {J.}~\bibnamefont {Ke}}, \bibinfo {author} {\bibfnamefont {W.-C.}\ \bibnamefont {Dong}}, \bibinfo {author} {\bibfnamefont {S.-H.}\ \bibnamefont {Huang}}, \bibinfo {author} {\bibfnamefont {Y.-J.}\ \bibnamefont {Tan}}, \bibinfo {author} {\bibfnamefont {W.-H.}\ \bibnamefont {Tan}}, \bibinfo {author} {\bibfnamefont {S.-Q.}\ \bibnamefont {Yang}}, \bibinfo {author} {\bibfnamefont {C.-G.}\ \bibnamefont {Shao}},\ and\ \bibinfo {author} {\bibfnamefont {J.}~\bibnamefont {Luo}},\ }\bibfield  {title} {\bibinfo {title} {{Electrostatic effect due to patch potentials between closely spaced surfaces}},\ }\href {https://doi.org/10.1103/PhysRevD.107.065009} {\bibfield  {journal} {\bibinfo  {journal} {Phys. Rev. D}\ }\textbf {\bibinfo {volume} {107}},\ \bibinfo {pages} {065009} (\bibinfo {year} {2023})}\BibitemShut {NoStop}%
\bibitem [{\citenamefont {Brownnutt}\ \emph {et~al.}(2015)\citenamefont {Brownnutt}, \citenamefont {Kumph}, \citenamefont {Rabl},\ and\ \citenamefont {Blatt}}]{IonTraps}%
  \BibitemOpen
  \bibfield  {author} {\bibinfo {author} {\bibfnamefont {M.}~\bibnamefont {Brownnutt}}, \bibinfo {author} {\bibfnamefont {M.}~\bibnamefont {Kumph}}, \bibinfo {author} {\bibfnamefont {P.}~\bibnamefont {Rabl}},\ and\ \bibinfo {author} {\bibfnamefont {R.}~\bibnamefont {Blatt}},\ }\bibfield  {title} {\bibinfo {title} {{Ion-trap measurements of electric-field noise near surfaces}},\ }\href {https://doi.org/10.1103/RevModPhys.87.1419} {\bibfield  {journal} {\bibinfo  {journal} {Rev. Mod. Phys.}\ }\textbf {\bibinfo {volume} {87}},\ \bibinfo {pages} {1419} (\bibinfo {year} {2015})}\BibitemShut {NoStop}%
\bibitem [{\citenamefont {Roberts}\ and\ \citenamefont {Ursell}(1960)}]{diffsphere}%
  \BibitemOpen
  \bibfield  {author} {\bibinfo {author} {\bibfnamefont {P.~H.}\ \bibnamefont {Roberts}}\ and\ \bibinfo {author} {\bibfnamefont {H.~D.}\ \bibnamefont {Ursell}},\ }\bibfield  {title} {\bibinfo {title} {{Random Walk on a Sphere and on a Riemannian Manifold}},\ }\href {http://www.jstor.org/stable/73126} {\bibfield  {journal} {\bibinfo  {journal} {Philosophical Transactions of the Royal Society of London. Series A, Mathematical and Physical Sciences}\ }\textbf {\bibinfo {volume} {252}},\ \bibinfo {pages} {317} (\bibinfo {year} {1960})}\BibitemShut {NoStop}%
\bibitem [{\citenamefont {Ledesma-Dur{\'a}n}\ and\ \citenamefont {Ju{\'a}rez-Valencia}(2023)}]{Ledesma}%
  \BibitemOpen
  \bibfield  {author} {\bibinfo {author} {\bibfnamefont {A.}~\bibnamefont {Ledesma-Dur{\'a}n}}\ and\ \bibinfo {author} {\bibfnamefont {L.~H.}\ \bibnamefont {Ju{\'a}rez-Valencia}},\ }\bibfield  {title} {\bibinfo {title} {{Diffusion coefficients and MSD measurements on curved membranes and porous media}},\ }\href {https://doi.org/10.1140/epje/s10189-023-00329-z} {\bibfield  {journal} {\bibinfo  {journal} {The European Physical Journal E}\ }\textbf {\bibinfo {volume} {46}},\ \bibinfo {pages} {70} (\bibinfo {year} {2023})}\BibitemShut {NoStop}%
\bibitem [{\citenamefont {Russano}\ \emph {et~al.}(2018)\citenamefont {Russano}, \citenamefont {Cavalleri}, \citenamefont {Cesarini}, \citenamefont {Dolesi}, \citenamefont {Ferroni}, \citenamefont {Gibert}, \citenamefont {Giusteri}, \citenamefont {Hueller}, \citenamefont {Liu}, \citenamefont {Pivato} \emph {et~al.}}]{Giuliana}%
  \BibitemOpen
  \bibfield  {author} {\bibinfo {author} {\bibfnamefont {G.}~\bibnamefont {Russano}}, \bibinfo {author} {\bibfnamefont {A.}~\bibnamefont {Cavalleri}}, \bibinfo {author} {\bibfnamefont {A.}~\bibnamefont {Cesarini}}, \bibinfo {author} {\bibfnamefont {R.}~\bibnamefont {Dolesi}}, \bibinfo {author} {\bibfnamefont {V.}~\bibnamefont {Ferroni}}, \bibinfo {author} {\bibfnamefont {F.}~\bibnamefont {Gibert}}, \bibinfo {author} {\bibfnamefont {R.}~\bibnamefont {Giusteri}}, \bibinfo {author} {\bibfnamefont {M.}~\bibnamefont {Hueller}}, \bibinfo {author} {\bibfnamefont {L.}~\bibnamefont {Liu}}, \bibinfo {author} {\bibfnamefont {P.}~\bibnamefont {Pivato}}, \emph {et~al.},\ }\bibfield  {title} {\bibinfo {title} {{Measuring fN force variations in the presence of constant nN forces: a torsion pendulum ground test of the LISA Pathfinder free-fall mode}},\ }\href {https://doi.org/10.1088/1361-6382/aaa00f} {\bibfield  {journal} {\bibinfo  {journal} {Classical and Quantum Gravity}\ }\textbf {\bibinfo {volume} {35}},\ \bibinfo
  {pages} {035017} (\bibinfo {year} {2018})}\BibitemShut {NoStop}%
\bibitem [{Note5()}]{Note5}%
  \BibitemOpen
  \bibinfo {note} {We use the one-sided ASD, defined as the square root of twice the Fourier transform of the autocorrelation.}\BibitemShut {Stop}%
\bibitem [{\citenamefont {Gelman}\ \emph {et~al.}(2020)\citenamefont {Gelman}, \citenamefont {Vehtari}, \citenamefont {Simpson}, \citenamefont {Margossian}, \citenamefont {Carpenter}, \citenamefont {Yao}, \citenamefont {Kennedy}, \citenamefont {Gabry}, \citenamefont {B\"{u}rkner},\ and\ \citenamefont {Modr\'{a}k}}]{gelman2020bayesian}%
  \BibitemOpen
  \bibfield  {author} {\bibinfo {author} {\bibfnamefont {A.}~\bibnamefont {Gelman}}, \bibinfo {author} {\bibfnamefont {A.}~\bibnamefont {Vehtari}}, \bibinfo {author} {\bibfnamefont {D.}~\bibnamefont {Simpson}}, \bibinfo {author} {\bibfnamefont {C.~C.}\ \bibnamefont {Margossian}}, \bibinfo {author} {\bibfnamefont {B.}~\bibnamefont {Carpenter}}, \bibinfo {author} {\bibfnamefont {Y.}~\bibnamefont {Yao}}, \bibinfo {author} {\bibfnamefont {L.}~\bibnamefont {Kennedy}}, \bibinfo {author} {\bibfnamefont {J.}~\bibnamefont {Gabry}}, \bibinfo {author} {\bibfnamefont {P.-C.}\ \bibnamefont {B\"{u}rkner}},\ and\ \bibinfo {author} {\bibfnamefont {M.}~\bibnamefont {Modr\'{a}k}},\ }\href {https://doi.org/10.48550/arXiv.2011.01808} {\bibinfo {title} {{Bayesian Workflow}}} (\bibinfo {year} {2020}),\ \Eprint {https://arxiv.org/abs/2011.01808} {arXiv:2011.01808 [stat.ME]} \BibitemShut {NoStop}%
\end{thebibliography}%
\end{document}